\documentclass[runningheads]{cl2emult}

\usepackage{makeidx}  
\usepackage{graphicx} 
\usepackage{subeqnar} 
\usepackage{multicol} 
\usepackage{cropmark} 
\usepackage{lnp}      
\makeindex            

\begin{document}
\setcounter{secnumdepth}{3}
\title*{Modelling Selforganization and
Innovation Processes in Networks}
\toctitle{Modelling Selforganization and
Innovation Processes in Networks}
%
%
\titlerunning{Modelling Selforganization and
Innovation Processes in Networks}
%
\author{Ingrid Hartmann-Sonntag\inst{1}
\and Andrea Scharnhorst\inst{2}
\and Werner Ebeling\inst{3}}
\authorrunning{Hartmann-Sonntag et al.}
%
%
\institute{Humboldt-Universit\"at zu Berlin, Institut f\"ur Physik,
Newtonstr.\ 15, 12489 Berlin, Federal Republic of Germany;
E-mail: {\tt uwe.hartmann@fh-stralsund.de}
\and Networked Research and Digital Information,
The Netherlands Institute for Scientific Information Services (NIWI),
Royal Netherlands Academy of Arts and Sciences,
Joan Muyskenweg 25, Postbus 95110, 1090 HC Amsterdam,
The Netherlands; E-mail: {\tt Andrea.Scharnhorst@niwi.knaw.nl}
\and Humboldt-Universit\"at zu Berlin, Institut f\"ur Physik,
Newtonstr.\ 15, 12489 Berlin, Federal Republic of Germany;
E-mail: {\tt ebeling@physik.hu-berlin.de}}

\maketitle              

\begin{abstract}
In this paper we develop a theory to describe innovation processes in a
network of interacting units. We introduce a stochastic picture that allows
for the clarification of the role of fluctuations for the survival of innovations 
in such a non-linear system. We refer to the theory of complex 
networks and introduce
the notion of sensitive networks. Sensitive networks are networks in which
the introduction or the removal of a node/vertex dramatically changes the dynamic
structure of the system. As an application we consider interaction networks
of firms and technologies and describe technological innovation as a
specific dynamic process. Random graph theory, percolation, master equation
formalism and the theory of birth and death processes are the mathematical
instruments used in this paper.
\end{abstract}

\section{Introduction}
This paper is devoted to the interdisciplinary theory of selforganization
processes, paying particular attention to stochastic effects connected with
innovations in network systems. In our understanding ``selforganization'' is
the spontaneous formation of structures \cite{ebel1982,ebel1994,feis1989}.
An ``innovation'', in a general system-theoretical understanding, is
the appearance of a new species, of a new mode of behavior, of a new
technology, of a new product or of a new idea etc.\ 
\cite{ebel1986bio,bruc1989,bruc1990,bruc1996b,ebel1999}.

Technological innovations are considered as the basic driving process
for economic evolution and growth. In economics, innovation networks
\cite{fren2000,kowo2003} and networks economies \cite{nagu2003}
have been discussed widely in the last decade. However, a
unified understanding of socio-economic networks is still missing 
\cite{savi2000net}. 
``For many economists the study of networks is limited to the
analysis of the functioning of physical networks such as the railway, the
telephone system or the internet for example'' 
\cite{kirm2003}.
Although networks are thought to be constituted by sets of actors and by links
\cite{savi2000net},
the very nature of these actors and the links between them
varies among different authors. Nations, institutions, firms, products, or
individuals may represent the actors. The links can be defined
quite differently. Innovation in such a context is mostly seen as an outcome
of complex networks with heterogeneous actors. 

In this paper, we define innovation not as a product 
of a network activity but as a specific process in the formation
of a complex network. Thus, innovation is a process in the 
structure formation of a complex networks. 
We describe innovation as a process
which dramatically and decisively changes the composition of a 
network. Accordingly, the network will behave differently depending
on which innovation has been introduced. More specifically, we propose 
a network description for dynamic processes in a system. 
By means of this system-theoretical perspective the appearance of innovations 
is related to certain mechanisms in the growth and change of networks. 
So, we create a network picture for dynamic processes as the emergence,
survival or extinction of innovations which have been also 
considered in the mathematical theory of biological evolution, in
population biology and other fields of complexity theory. 

In this paper we will follow a special approach to describe innovations in
evolutionary systems. We start in section 1 with an abstract definition of
an innovation. We will then show that our approach is related to recent
developments in statistical physics which best can be described as an
emerging field of complex networks theory 
\cite{scha2003}.
Thus we will first give a survey of network
approaches ranging from chemical networks 
\cite{temk1996,fell1997} to biochemical webs \cite{fell1997}, 
protein webs \cite{jeon2000}, food webs \cite{dros2003},
to the structure of the internet \cite{falo1999}
and the world wide web \cite{albe1999}
up to general approaches from statistical mechanics 
\cite{albe2002}. 

In the second part of the paper we reflect upon
structure and relations of socio-economic networks in a static
(purely structural) picture. We will discuss some earlier results
of random graph theory and percolation theory. We derive statements about
connectivity and strongly connected components.

In the third part of the paper, and in connection with our special interest
in innovation processes, we will develop a dynamic network theory
and in particular the theory of ``sensitive'' networks in more detail.
The term ``sensitive'' networks denotes
networks which are sensitive to the introduction or removal of one or few
nodes or edges, or in a more general context, to the occupation or leaving
of a node. Specifically, sensitivity  is linked to the question of whether
a node (a species, a mode of behavior, an idea, a technology) is occupied by
at least one individual or not. We will show that this problem is of
relevance for the modeling of innovation processes. The dynamics of
innovation processes may be described by stochastic equations, which can be
formally treated within the framework of statistical physics. Some
generalizations are possible.

\subsection{\label{sec1.1}Innovation -- a System Theoretical Approach}
In economics, innovation is mainly understood as technological innovation
describing the introduction of new technologies, products and production
processes. The differentiation between invention and innovation relates
innovation to the economic exploitation of new ideas. However, it is also
possible to look at innovations from a more general, evolutionary point of
view. Ziman gives one example for such an approach when he writes ``Go to a
technological museum and look at the bicycles. Then go to a museum of
archaeology and look at the prehistoric stone axes. Finally, go to a natural
history museum and look at fossil horses. In each case, you will see a
sequence, ordered in time of changing but somewhat similar objects.''
\cite{zima2000}, p.\ 3.
However, not every change is an innovation. In this
paper, we follow a system theoretical approach to innovation. In this framework,
innovation is something new to the system and most essentially 
the emergence of an innovation changes the state of the
system dramatically. In other words, the actual state of the system becomes unstable and
a transition to a new state occurs. To define an innovation we first have to
define the state of the system. Here, we again choose a very specific
approach. We represent the state of the system as a point in the high
dimensional occupation number space 
\cite{ebel1982,ebel1986bio}.
In this space a coordinate axis is attached to a certain type $i$
of elements (with $i=1,2,\ldots ,s$, natural numbers). 
The occupation numbers are represented on this axis.

To describe technological innovation we have to ask for a re-specification
of this abstract concept. For socio-economic systems, the axes of the state
space refer to different possible taxonomies. For instance, an axis $i$ can
represent a certain technology from a set $s$, of different technologies
present in the system. With such a technological taxonomy, competition
processes between technologies can be described \cite{savi1995,bruc1996b}.
The carriers of this competition process are firms
using different technologies and competing with their products on a market.
This way, we link back to an economic understanding of an innovation process
that ``requires insight into system dynamics grounded in a variety of firm
competencies and behaviour and a variety of demand'' 
\cite{savi2000noot}, p.\ 5.
Let us note that the state space concept can be applied to
quite different processes. The type $i$ might also stand for the size class
that a certain firm belongs to. Then, growth processes of firms are in the
focus of the description. Moreover, the type $i$ may represent a certain group
in society. Formation of political opinions \cite{weid2000}
or emergence of norms and violence in groups 
\cite{nach1998} are then considered.
Innovation in these cases covers new forms of collective behavior.

We can find each type $i$ in $N_i$ exemplars of elements in the system. 
The exemplars may be
individuals, but also organizational and institutional units like firms and
groups. $N_i$, the occupation numbers, are functions of time. They are
positive or zero. A complete set of occupation numbers 
$N_1,N_2,\ldots,N_s$ at a fixed time, characterizes the 
occupation state of this system.
The time dependent change of the occupation numbers is described by the
movement of this point in the space. The whole motion takes place on the
non-negative cone $K$ of the space. In this picture we can describe the case,
that a type $i$ is not present in the system at time $t$. 
That means the type $i$ is
occupied with the number zero ($N_i=0$). We call a system an
under-occupied system if we can make the assumption that the sum of the
occupation numbers is essentially smaller than the total number of the
possible elements \cite{ebel1986bio}.

In this picture, an innovation is an occupation of a non-occupied type. In
deterministic systems, zero occupation can only be achieved
in the limit of infinite time $t\to\infty $ , if a sort died out
(zero can be a stable stationary state). For finite times $t>0$ 
types cannot arise, if they are not in the system at time
$t=0$ , and present types cannot die out. The situation is
different if we use the stochastic picture. In 
stochastic systems the zero state
can be reached in finite times $t$. A stochastic description
offers the advantage that at finite times new sorts (innovations) can arise
or die out. Therefore, the stochastic description is especially suited for
evolutionary processes and in particular for innovation processes.

Let us note here that any innovation will change the 
taxonomy of types in the system \cite{alle1994}.
Innovation has to do with uncertainty and its prediction is
impossible. With the notion of an under-occupied system we escape the
problem of determining \textit{a priori} the place or kind of an 
innovation. Instead, we equip the system
with a reservoir of possible innovations. Which of these possibilities
turns into a realization remains uncertain. In some respects this is a trick
to avoid the problem with a changing taxonomy. There are other possibilities
to escape this problem, e.g., so-called continuous models operating on a
characteristic space as we discussed elsewhere 
\cite{ebel1999,ebel2000traffic,ebel2001}.
However, the discrete approach has, in our respect, certain
advantages, as we will discuss later.

In an under-occupied system most elements have, at a given time $t$, the
occupation number zero. So we can pass from the high-dimensional cone $K$
to a lower-dimensional cone $K^+$. Accordingly,
the time dependent variation of the system can be described as a switching
of the state point on the edges of the cone $K$. If $K^+$ is an
element of the set of all possible cones, we observe 
a switch from one sub-cone to another. As the process is 
discrete, it is a hopping on the edges of different positive cones.

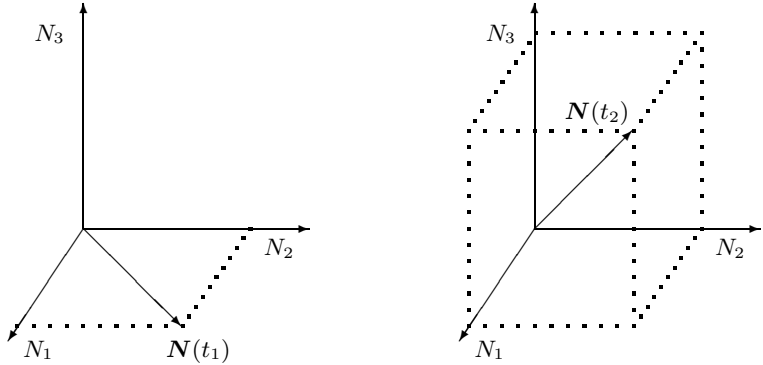
\begin{figure}[h]
\unitlength1.mm
\begin{picture}(139,55)
\put(7.5,0){
\begin{picture}(50,50)
\linethickness{0.15mm}
\put(10,25){\vector(1,0){30}}
\put(10,25){\vector(0,1){30}}
\put(10,25){\vector(-2,-3){10}}
\linethickness{0.4mm}
\bezier{10}(32,25)(27.5,18.5)(23,12)
\bezier{10}(1,12)(12,12)(23,12)
\linethickness{1mm}
\put(10,25){\vector(1,-1){13}}
\put(2,8){$N_1$}
\put(34,21.5){$N_2$}
\put(3.5,50){$N_3$}
\put(21,8){$\vec{N}(t_1)$}
\end{picture}  }
\put(67.5,0){
\begin{picture}(50,50)
\linethickness{0.15mm}
\put(10,25){\vector(1,0){30}}
\put(10,25){\vector(0,1){30}}
\put(10,25){\vector(-2,-3){10}}
\linethickness{0.4mm}
\bezier{10}(32,25)(27.5,18.5)(23,12)
\bezier{10}(1,12)(12,12)(23,12)
\bezier{10}(32,51)(27.5,44.5)(23,38)
\bezier{10}(1,38)(12,38)(23,38)
\bezier{10}(10,51)(5.5,44.5)(1,38)
\bezier{10}(10,51)(21,51)(32,51)
\bezier{12}(1,12)(1,25)(1,38)
\bezier{12}(23,12)(23,25)(23,38)
\bezier{12}(32,25)(32,38)(32,51)
\linethickness{1mm}
\put(10,25){\vector(1,1){13}}
\put(2,8){$N_1$}
\put(34,21.5){$N_2$}
\put(3.5,50){$N_3$}
\put(14,39.5){$\vec{N}(t_2)$}
\end{picture}  }
\end{picture}
\caption[]{The occupation number space: The distribution of individuals
over different types is
represented by a vector in the positive cone. The appearance or 
extinction of a type
can be described as approaching or leaving one of the edges 
of the positive cone.}
\label{fig1}
\end{figure}

In figure \ref{fig1} 
we visualize such a process for three dimensions. At any point in time,
the state of the system is represented by a certain vector 
$\vec{N}(t)$. In a stationary state, the endpoint of this vector 
defines a positive cone. In our example, the vector moves 
in the plane spanned by $N_1$ and $N_2$. 
An innovation opens up a new dimension of the system. 
In our example, a new third type is introduced into the system.
After the innovation, the vector is moving in the space defined by $N_1$,
$N_2$ and $N_3$. In general, we can assume that the system operates 
in a multidimensional space where the cone can have a very complicated 
shape, and the vector $\vec{N}(t)$  
jumps between the edges of this cone.

The hopping process visualizes the transition between one stable stationary
state and another stable stationary state. In this sense, innovation is the
outcome of a process of destabilization. In the framework we propose in this
paper, innovations are seen as stochastic instabilities. The changes
occurring in the occupation number space result from interactions of the
different types present in the system. These interactions can be visualized
as graphs or networks where the nodes represent the types, and links between
them represent different forms of interactions. In the network picture, an
innovation corresponds to the appearance of a new node and the activation
of a link to this node. The models we will
present in section 3 allow us to differentiate between different processes
which finally introduce such a new node.

The conceptualization of types as elements (nodes) of a network
represents a graph theoretical approach to the dynamics of the system.
Therefore, other network approaches are of particular relevance to develop our
theoretical approach further. We will use the following part of section 1 to
introduce the new emerging specialty of complex network theory and to place
our approach in this field.

\subsection{Innovations, the Emergence of the Field of Complex Network 
Theory and Sensitive Networks}
In the last years, complex systems in nature and society have been carefully
investigated. Already in the 70s, theories of self-organization
were used to build a bridge between social and natural systems
investigations \cite{prig1987}. As part of this
development complex networks have been investigated.
Recently, as a new branch in complexity theory \cite{schw1997self}
complex networks have been reconsidered and extensively studied 
\cite{scha2003}. They seem to be particularly relevant for the study of
innovation processes. 

In the context of complexity theory, the concept of networks has not only been used
as an easy-to-use metaphor. As Bornholdt and Schuster note:
``Recent advances in the theory of complex networks indicate that this
notion may be more than just a philosophical term. Triggered by recently
available data on large real world networks (e.g., on the structure of the
internet or on the molecular networks in the living cell) combined with fast
computer power on the scientist's desktop, an avalanche of quantitative
research on network structure and dynamics currently stimulates diverse
scientific fields.''\cite{born2003}, p.\ V.
Social networks form one important area of application of complex networks
theory. The structures found cover networks of collaboration 
\cite{newm2001,bara2002phys}, 
networks of recognition (citation networks) \cite{vazq2001}  
and networks of corporate directors \cite{stro2001}.
In economic theories, innovation is more and more
understood as the outcome of the interaction between scientific, economic
and political systems \cite{pyka2002}.
Instead of considering an
innovation as a singular event the network character of innovations is
stressed. Innovation networks seem to be a new organizational form of
knowledge production.

The structural analysis of systems is of great interest. Albert and
Barab\'asi \cite{albe2002} 
give a very good presentation of this subject and its
development. In the very beginning, investigations of large complex systems
were done by random graphs. More and more it became possible to analyze real
complex systems and large systems too. With the development of computer
capacity, the amount of empirical data increases. It becomes possible to
compare the theoretical results by random graph theory with that of the real
data analysis. Obviously more than pure randomness exists. Organization
principles and rules of system evolution play a decisive role, leading to
small-world behaviour and scale-free networks. Our world is not a random
world. Other evolutionary principles are of great interest. In addition, it
is evident that a theory of evolving networks may give a more realistic
approach to real systems. This is why we give special
consideration to evolving networks here.

 From the analysis of empirical data we learn that many real-networks have a
small-world character. The small-world concept describes the fact that
despite their often large size, in most networks there is a relatively short
path between any two nodes. The small-world property characterizes
most complex networks. For example, the chemicals in a cell are typically
separated by only three reactions, or in a more exotic case, the actors in
Hollywood are on average within three co-stars from each other. The
small-world concept corresponds to our observations; it is a structural, not
an organizing, principle \cite{watt1999,buch2002}.

Not all nodes in a network have the same number of edges (the same node
degree). The spread in node degrees is characterized by a distribution
function $P(k)$, which gives the probability that a randomly selected node has
exactly $k$ edges. Since in a random graph the edges are placed randomly, the
majority of nodes have approximately the same degree, close to the average
degree of the network. The degree-distribution of a random graph is a
Poisson distribution with the peak over the average degree 
\cite{albe2002}. In real networks the distributions of the edges
are more complicated. Important results were obtained by the analysis of
large real systems: the degree distribution deviates significantly from a
Poisson distribution and follows general structural rules in many cases.
Many large networks are scale-free, that is, their degree distribution
follows a power law. In addition, even for those networks for which $P(k)$ has
an exponential character, the degree distribution significantly deviates
 from a Poisson distribution achieved by random graph theory for such systems
\cite{albe2002,bara2002new}.

Scale-free networks express a hierarchy between the nodes. Not every node is
important at the same level. Accordingly, not each of the links between the
types have the same importance. There are very sensitive relations or
elements too. As mentioned previously, we consider networks as ``sensitive''
if their properties depend strongly on the introduction or removal of one or
a few nodes or edges, or on changes of the occupation of nodes. 
We will show that for the
evolutionary character, the description of the time-behaviour by master
equations on occupation number spaces is an appropriate tool. The discrete
character of occupation number description allows for an appropriate
description of the introduction respectively of removal of relations, edges
etc.. We will analyze not only the steady states of our stochastic systems
but also the time evolution. Albert and Barab\'asi \cite{albe2002}
also refer to
approaches with master and rate equations. They write that in addition, these
methods, not using a continuum assumption, appear more suitable for
obtaining exact results in more challenging network models. In addition they
mentioned that the functional form of the degree distribution, $P(k)$, cannot be
guessed until the microscopic details of the network evolution are fully
understood. According to our point of view, the method of master equations is
an excellent tool to use in the investigation of many open questions and is able
to bring much more light to bear on this subject. For example, by using this
discrete approach, we have the chance to get statements about the kind of
fluctuations. Evidently this is one of the most important questions.

Let us come back now to the question of the distribution of the graph. We
remember that random graph theory leads to a Poisson degree-distribution.
Albert and Barab\'asi \cite{albe2002} 
give a near exhaustive survey of empirical
data sets for real complex networks and show that the real
degree-distributions are not Poisson-distributions, but
scale-free-distributions, or exponential distributions. These authors write:
``The high interest in scale-free networks might give the impression that
all complex networks in nature have power-law degree distributions. \ldots
It is true that several complex networks of high interest for scientific
community, such as the world wide web, cell networks, the internet, some
social networks, and the citation network are scale-free. However, others,
such as the power grid or the neural networks of c.\ elegans, appear to be
exponential \ldots Evolving networks can develop both power-law and
exponential degree distributions. While the power-law regime appears to be
robust, sublinear preferential attachment, aging effects, and growth
constraints lead to crossovers to exponential decay. \ldots If all
processes shaping the topology of a certain network are properly
incorporated, the resulting $P(k)$ often has a rather complex form, described
by a combination of power laws and exponentials.``

In this respect our aim here is the calculation of the role of fluctuations
by the use of the master equation approach. This way we can make statements
as to how the systems differ from linear systems which obey a
Poisson-distribution. In principle, by investigating the fluctuations, correlations and
spectral-densities, we are able to study several microscopic events. One of
the questions to solve is which fluctuation effects produce power-law
distributions. We see a deep connection of these network systems to systems
which produce $1/f$-noise. We remember that $1/f$-noise is a stochastic
process with a specific power-law spectrum \cite{klim1995}.
A characteristic property of processes which produce $1/f$-noise are long
range-correlations. We suppose that the scale-free networks and
small-world-behaviour may have some relation to this. We remember that in
small-world networks the degree-function obeys an power-law; there exists a
small pathway between each two of the elements.

In investigating selforganization and evolution processes in networks,
our basic approach is that we understand the corresponding
networks as dynamic, or more precisely, as evolutionary systems. This
dynamic and evolutionary approach allows us to make statements about
innovation processes, special competition effects, the sensitivity of
networks, the constraints of growth and the fitness of network systems.

\section{Structure and Relations of Socio-Economic Networks}

\subsection{Structure, Selforganization and Complexity}
With respect to their structure, social connections are relations between
elements. Therefore, as can be found in each handbook of social network
analysis, relational data form the basis of social networks 
\cite{scot2000,wass1994}. Such data describe ties, connections, group
attachments, meetings and other events or activities that relate one
individual to another one. From the abstract point of view, socio-economic
networks are ``structures''. It is possible to give an abstract presentation
of such structures by means of mathematical tools. The mathematical idea of
structure stands in close connection to the terms

\vspace{0.3cm}

\hspace{1cm}element, set, relation and operation.

\vspace{0.3cm}

\noindent
The nature of the elements does not play any role with respect to the
structure. Opposite to this, the nature of the relations between elements
determines the specificity of a structure. Structure here means the
manifold of interactions between the elements.

Let us note here, that for many descriptions of socio-economic networks as
in sociology and economics, it is just the nature of the elements and the
relations which are relevant. In this paper, we follow a more abstract and
general approach. By using such a general level, we create methodologically
the task of re-specifying the definition of both elements and relations
for any application area that one might think of. On the other hand, the high
level of generalization we use here keeps the application areas open, as we
intend to do.

The idea of ``structure'' is of great importance in our life -- both in
reality and in sciences. In particular, it is obvious by analyzing social
structures. What does ``structure'' mean in the original sense? On the one
hand we have our conventional understanding of structure -- the understanding of
structure in our real life. On the other we may formulate a precise
idea of structure in terms of mathematics, in system theory 
\cite{cast1979} and in the theory of selforganization \cite{ebel1998}:

\vspace{0.3cm}

\noindent
``We understand under a ``structure'' the composition of elements and the
set of relations respectively operations, which connect the elements.''
\vspace{0.3cm}

\noindent
Kr\"ober \cite{krob1967}
writes about the idea of structure in real systems: ``Each
system consists of elements that are arranged in a certain way and are
linked to each other by relations. We understand by 'structure of a system'
the kind of arrangement and connections of their elements \ldots 
In this respect, we do not consider what kind the
elements are. If we speak about
structure, we are not interested in the elements of the structure. We only
consider the manifold of relations. In this respect the structure of a
system is a well-defined connection between the elements of the system.
These elements, which are arranged in a determined manner and connected by
determined connections can be necessary or randomly, universally or uniquely,
relevant or irrelevant''.

Moles \cite{mole1962}
writes to this subject: ``The surrounding objects of the
material world, artifical and natural organism in the wide sense of this
world are signed by two main aspects; by their structural and functional
properties.''

With the famous book series ``The Elements of Mathematics'' the group of
scientists, Bourbaki, gave an example for constructing systematically
mathematics as a science of such ``structures''. In the following section we
give a short description of several important
concepts of this mathematical theory of structures. In particular, we
give a summary of the theory of relation, graphs and matrices in the
amount we need here. Our purpose is to apply this abstract theory to
socio-economic networks.

In a socio-economic system, the elements of
structure are individuals or groups of individuals in different
institutional and organizational forms, e.g., firms. The socio-economic
connections between these elements are relations in the sense of this
abstract theory of structure. The description of elements and relations 
can be given graphically by
a system of vertices (nodes), which model the elements (individuals, groups, firms)
and of edges (arcs), which describe the relations (connections). Nodes and arcs
can be weighted. Arcs can have a direction. This way we can include
quantitative aspects. The most important aspect in the selforganization of networks is
the formation of new connections, which generates new structures. In the
following, we want to show that the instruments of the mathematical theory of
structure in connection with the ideas and concepts of the theory of
selforganization contribute to the description of the connections in
complex socio-economic networks.

The idea of ``structure'' stands in a close connection to order (disorder)
respectively information (entropy). The theory of selforganization shows
how the generation of structures is connected with the decrease of entropy
\cite{ebel1998}. 

Socio-economic systems are complex systems, which consist of
many connections between the elements. Therefore, complexity is a further
concept to be defined. Ebeling, Freund and Schweitzer 
\cite{ebel1998}
write:
\vspace{0.3cm}

\noindent
``As complex we describe holistic structures consisting of many components,
which are connected by many (hierarchically ordered) relations respectively
operations. The complexity of a structure can be seen in the number of equal
respectively distinct elements, in the number of equal respectively distinct
relations and operations, as well as in the number of hierarchical levels. In
the stricter sense, complexity requires that the number of elements becomes
very large (practically infinite).''
\vspace{0.3cm}

\noindent
We are especially interested here in the origin of complex structures, and 
 in the development of order (information). 
In the end we have to answer the question
which parameter-relations (order parameters) determine the qualitative
behaviour of the system. Ilya Prigogine \cite{prig1955} 
in collaboration with his coworkers, did pioneering work in the investigation of
selforganizing systems. \cite{nico1977}
Further important work has been done in this
field by Manfred Eigen \cite{eige1971} 
on the selforganization of
macromolecules and by Manfred Eigen and Peter Schuster 
\cite{eige1971,eige1978c} 
on the hypercycle model. The mechanisms of selforganization are
clearly worked out by Nicolis and Prigogine 
\cite{nico1977}.  These authors give a
stringent physical and mathematical formulation for these processes, in
particular with respect to the energetic and entropic aspects. A somewhat
different view on this was developed in the formulation of the synergetics
by Haken \cite{hake1978}. 
The investigation of such systems shows that the formation
of order in complex systems can be allocated to physical processes which
play a role far from equilibrium \cite{ebel1982}. 
We underline that biological, just as socio-economic processes, can be
investigated with the help of the theory of selforganization because they
obey the valid physical and chemical laws. However, processes which include
the real life (biological and socio-economic systems) also obey other rules
and laws, which are not solely determined by physics. This is already evident
 from the very general character of the structures we consider here. As said
above, we formulate the idea of structure mathematically and keep away from
how the structure appears in different systems. Then, graph theory will provide us 
with the level of abstraction we need in
order to describe real systems.

\subsection{Introduction to the Theory of Relations, Matrices and Graphs}

By formulating our ideas in the mathematical language, we have the advantage
of having access to the great mathematical potential which is available in
this field. An important basic assumption is to start from the theory of sets.
As pointed out above, the connection between element and set is the first
and most important aspect of a structure. Furthermore, we introduce relations
and operations. The concept of structure reflects abstract properties of a
system. Due to the abstract character of the concept, results can be
translated to other systems and comparisons are possible. 

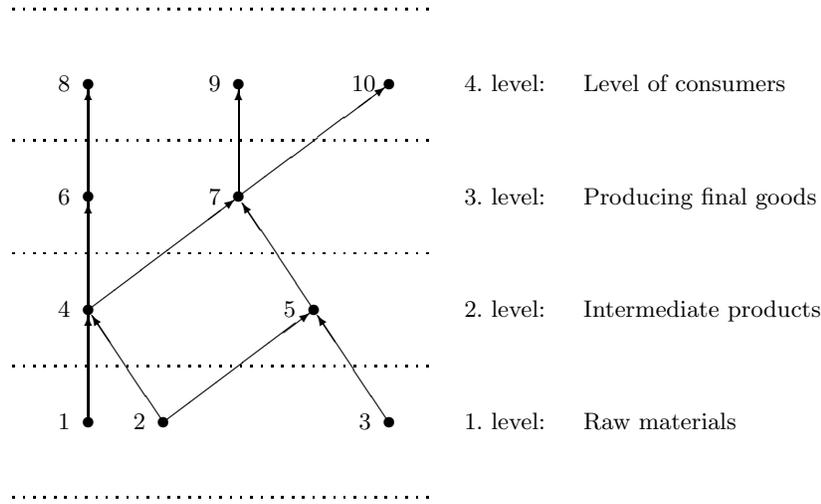
\begin{figure}[h]
\unitlength1.mm
\begin{center}
\begin{picture}(110,80)
\put(10,10){\circle*{1.5}}
\put(10,10){\vector(0,1){14.25}}
\put(20,10){\circle*{1.5}}
\put(20,10){\vector(-2,3){9.5}}
\put(20,10){\vector(4,3){19.5}}
\put(50,10){\circle*{1.5}}
\put(50,10){\vector(-2,3){9.5}}
\put(10,25){\circle*{1.5}}
\put(10,25){\vector(4,3){19.5}}
\put(10,25){\vector(0,1){14.25}}
\put(40,25){\circle*{1.5}}
\put(40,25){\vector(-2,3){9.5}}
\put(10,40){\circle*{1.5}}
\put(10,40){\vector(0,1){14.25}}
\put(30,40){\circle*{1.5}}
\put(30,40){\vector(0,1){14.25}}
\put(30,40){\vector(4,3){19.5}}
\put(10,55){\circle*{1.5}}
\put(30,55){\circle*{1.5}}
\put(50,55){\circle*{1.5}}
\linethickness{0.2mm}
\bezier{40}(0,0)(27.5,0)(55,0)
\bezier{40}(0,17.5)(27.5,17.5)(55,17.5)
\bezier{40}(0,32.5)(27.5,32.5)(55,32.5)
\bezier{40}(0,47.5)(27.5,47.5)(55,47.5)
\bezier{40}(0,65)(27.5,65)(55,65)
\put(6,9){$1$}
\put(6,24){$4$}
\put(6,39){$6$}
\put(6,54){$8$}
\put(16,9){$2$}
\put(46,9){$3$}
\put(36,24){$5$}
\put(26,39){$7$}
\put(26,54){$9$}
\put(45,54){$10$}
\put(60,9){\rm 1.\ level:\hspace{.5cm}Raw materials}
\put(60,24){\rm 2.\ level:\hspace{.5cm}Intermediate products}
\put(60,39){\rm 3.\ level:\hspace{.5cm}Producing final goods}
\put(60,54){\rm 4.\ level:\hspace{.5cm}Level of consumers}
\end{picture}
\end{center}
\caption[]{Graph of an economic network with 4 levels: 
(i) raw material such as minerals,
fossils, plants etc., (ii) intermediate products as steel, coal, corn, etc.,
(iii) producing final goods, (iv) level of consumers}
\label{fig2}
\end{figure}

In general we can
differ between local, temporal, causal and functional structures.
To illustrate the structures we will use graphs which represent the elements
and their connections by geometrical symbols 
\cite{hara1965,laue1970,cast1979,ebel1982,feis1989}.
An example
of a graph representing an economic network with 4 levels is given in 
Fig.\ \ref{fig2}. This graph represents the flow of materials and outcomes 
in a production process.

In economics, the relations between economic agents can be represented
in a network form. In this case information flows, e.g., price signals
between market participants, are exchanged. Then, the structure of the network
describes a situation with local interaction (not every agent is
informed about all other agents but the agents also do not act independently)
\cite{kirm2003}. Socio-economic networks can also describe a variety of
different agents' actions that influence other agents. The diffusion of
technologies over firms can be described as a network of actions from
formation of a company (with a certain technology) over knowledge transfer
between companies (in form of imitation or merging) to the exit of companies
due to technological competition. We will come back to such a network
interpretation in section 3.

It may be worthwhile to underline that it is not possible to give a complete
structural description of living objects by binary relations as
graphs only. Graphs are just a device for analyzing these systems,
albeit a very useful tool. Of course, we cannot describe complex objects 
only by graphs because their binary relations are ambiguous \cite{rash1965}. 
Nevertheless, graphs are very useful for the representation of complex
structures. Before the elements of the theory of graphs are explained 
it is necessary
to introduce the two notions; \textit{set} and \textit{relation}.

\begin{figure}[h]
\centering
\unitlength1.mm
\begin{picture}(110,15)
\put(10,10){\circle*{1.5}}
\put(9,5){{\large $a_1$}}
\put(30,10){\circle*{1.5}}
\put(29,5){{\large $a_2$}}
\put(50,10){\circle*{1.5}}
\put(49,5){{\large $a_3$}}
\put(70,8.5){{\large $M = \{a_1,a_2,a_3\}$}
}
\end{picture}
\caption[]{A set of three elements (for instance, plants or firms)
which are represented in a one dimensional space.}
\label{fig3}
\end{figure}
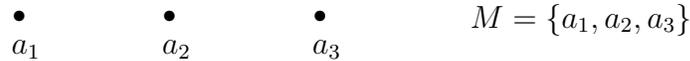

What we understand by a \textit{set} is nearly the same as in common 
language. It is always abstract and determined by its elements.
A set $M$ is given, which consists of the elements $a_1, a_2, a_3 \cdots$;
symbolically $M=\left\{a_1,a_2,a_3 \cdots \right\}$.The
elements $a_i$ ($i=1,2,3,\cdots$) describe individuals or groups
equipped with different attributes and features and belonging to
different types.
The number of elements determines whether we have a finite or infinite set.
Our investigations here are related to finite sets. In the 
following ``sets'' means finite sets. Graphically the elements of a set
can be described by vertices (nodes) in a n-dimensional space. 
This way we assume that a clear relation exists between the 
vertices and the elements of the set. An example is given in 
Fig.\ \ref{fig3}.

\noindent
The elements of a set can be ordered to pairs. Let us consider for example a
set $M = \{a_1,a_2,\ldots,a_5\}$ with the pairs $[a_1,a_3]$ and 
$[a_2,a_5]$. $a_1$ and $a_2$ describe the first
element and $a_3$ and $a_5$ the second one. The pair
$[a_5,a_2]$ describes another relation between the elements $a_5$ and 
$a_2$ as $[a_2,a_5]$. In general the
elements $a_i$ and $a_j$ in an ordered pair
$[a_i,a_j]$ are not exchangeable with each other without changing
the kind relations of the elements $a_i$ and $a_j$ to each other.

\begin{figure}[h]
\centering
\unitlength1.mm
\begin{picture}(100,25)
\put(30,20){\circle*{1.5}}
\put(29,21){{\large $a_1$}}
\put(20,10){\circle*{1.5}}
\put(19,5){{\large $a_2$}}
\put(40,10){\circle*{1.5}}
\put(39,5){{\large $a_3$}}
\put(20,10){\vector(1,0){19.25}}
\put(30,20){\vector(-1,-1){9.25}}
\put(30,20){\vector(1,-1){9.25}}
\put(60,8.5){{\large $M = \{a_1,a_2,a_3\}$}
}
\end{picture}
\caption[]{Directed graph}
\label{fig4}
\end{figure}
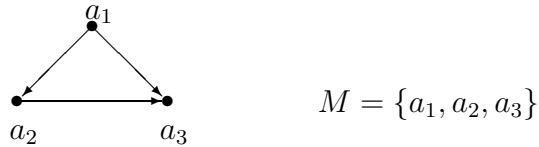

As an example we consider a set of firms with a certain economic structure
between them. The structure can be determined either in the form of trade
relations, financial transactions or information flows. Usually, relations
in an economy are directed. The set $M$ consists of three plants $a_1$,
$a_2$ and $a_3$. There are given
the following relationships: $a_1$ provides certain
products to $a_2$ and to $a_3$; $a_2$ delivers products to 
$a_3$. The ordered
pairs for these relation follows to:$[a_1,a_2]$, 
$[a_1,a_3]$ and $[a_2,a_3]$. We can represent the relationship
geometrically. This representation is called a directed graph
(Fig.\ \ref{fig4}).

\noindent
A graph consists of the elements of the set and the relations. In such a way
a graph can give determined relationships between elements of a set. In
general a graph is a structure model which mirrors aspects of 
the structure of the investigated object. In order to obtain 
this structure we have to consider
the relations between the corresponding partial objects. Let us note,
that both the notion "vertex" (vertices) and "node"
is used to describe the elements. Their relations are represented 
by "edges" or "arcs". Sometimes, the notion of an "arc" 
is used for an directed arc only. Other synonyms are link or line. 

We consider a system to be a certain representation of a real
phenomenon where certain boundaries between the system and its
environment can be defined. A
system is, in general, defined by the set of its elements and the set of
relations between the elements. Obviously, with regard to the original
phenomenon, different degrees of adjusting the model to the origin
are possible. If a unique map exists between the structure of the
original phenomenon and the structure of the model, the model is
called homeomorph. The border line case of
maximum adjustment of the model to the original in respect to the structure
is called isomorph.

\subsection{Structures with Random Relations and Their Application to
Socio-Economic Groups}
In the previous subsection we represented the individual units 
(e.g., plants or firms) as elements
(vertices) of a graph. The relation of the units are shown by
the edges. Let us introduce several simplifications. First we exclude the
relation of a node to itself (self-reflexibility) from our consideration, as this
leads to a special type of graph. Further we only consider graphs with
distinguishable vertices and indistinguishable directed edges. 
A directed graph is also called a digraph. We exclude loops
and parallel edges. We concentrate in this section on random graphs. The
capacity of a network (graph) -- the number of its elements -- is one of its
important functions. Of relevance for the structure is also the connectivity
of a network. Already 1970 Gardner and Ashby 
\cite{gard1970} investigated the probability of
the stability of great networks, in dependence on capacity and connectivity.
May \cite{may1972} performed similar investigations. Computer simulations
and some statements of probability about the structural behaviour
of networks may be found also in papers of Sonntag, Feistel,
Ebeling \cite{sonn1981} and Sonntag \cite{sonn1984}.
We now introduce several important terms that
characterize a network, following in part the work mentioned above.

If $K$ is the number of edges in the graph, and $S$ is the number of 
vertices, we denote the considered graph by $D(S,K)$. As a simple and 
important measure, the connectivity is defined as:
\begin{equation}
C\ =\ \frac{K}{S}\ .
\end{equation}

Isolated, connected parts of the digraph $D(S,K)$ consisting of
$s$ vertices and $k$ arcs are called components $d(s,k)$. Let us note
here that the ``basic step in the structural description of a network is to
identify the number and size of its components''
\cite{scot2000}, p.\ 102. 
Socio-economically interpreted, the existence, number and size of components
in a graph stands for the opportunities and obstacles to communicate, to
exchange information or/and to interact.

$S_k^s$ is the number of components with $s$ vertices and $k$ arcs. So
we can write:
\begin{equation}
\sum_{s,k}\ s\ S_k^s\ =\ S;\ \ \ \sum_{s,k}\ k\ S_k^s\ =\ K
\end{equation}
We denote the mean values of the frequency of the component $d(s,k)$ by:
\begin{equation}
H_k^s\ =\ \langle S_k^s\rangle\ =\ \sum_r\ r\ P_{k,r}^s,
\end{equation}
with $ P_{k,r}^s$ being the probability that in a special digraph
$D(S,K)$ $S_k^s = r$ components $d(s,k)$ can
occur.

As mean number of components of the digraph  $D(S,K)$ we use:
\begin{equation}
\eta\ =\ \sum_{s}\sum_k\ H_k^s .
\end{equation}
Apart from the trivial components, such as single vertices ($d(1,0)$) and
single arcs ($d(2,1)$), the digraph contains a ``structured
part'', whose number of components is written as follows:
\begin{equation}
L\ = \eta\ -\ H_0^1\ -\ H_1^2 .
\end{equation}
The probability distribution $P_{k,r}^s$ is not known for the finite
digraphs. In the limit $S\rightarrow\infty$ the probabilities for the
emergence of different finite large components are uncorrelated, so 
we have to work with a Poisson distribution for the possibility of $r$
components with $s$ vertices and $k$ arcs in the whole graph 
\cite{sonn1981}:
\begin{equation}
\lim_{S\rightarrow\infty}\ P_{k,r}^s\ =\
\frac{\left( H_k^s \right)^r}{r!}\ \exp\left(-H_k^s\right)
\ \ {\rm for}\ \ \ S\gg s .
\end{equation}

\begin{figure}
\vspace{2.5cm}
\centering
\includegraphics[width=.9\textwidth,angle=0]{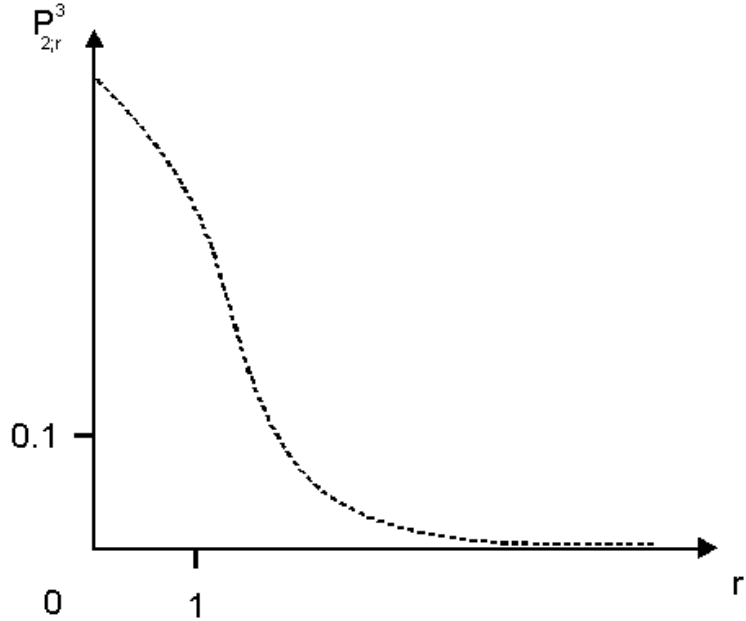}
\caption[]{The probability distribution $P_{2,r}^3$ for $C=0.05$ 
that $r$ components $d(3,2)$ occur}
\label{fig5}
\end{figure}

In dependence on $C$ we can for instance represent 
$P_{0,r}^1$, $P_{1,r}^2$, $P_{2,r}^3$, $P_{3,r}^3$ over $r$.
So we can show the building of components in dependence on the connectivity
$C$ (Fig.\ \ref{fig5}).

Following Sonntag, Feistel, Ebeling \cite{sonn1981}
we can achieve further statements.
In the limit $S\rightarrow\infty$ the mean number of components $\eta$ 
in practice is determined only by semicycleless components
(trees). So we can use the approximation:
\begin{equation}
\eta\ =\ \sum_{s=1}^S\ H_{s-1}^s\ +\ O\left(S^0\right)
\end{equation}
In that limit $S\rightarrow\infty$ the number of non-trivial components:
\begin{equation}
L\ =\ \eta\ -\ S\ \exp(-2C)\ -\ SC\ \exp(-4C)
\end{equation}
follows.

By a constant number of $S$ of individuals, firms (vertices) the building of
relations between these individuals, firms corresponds to an increasing
number $K$. $K$ is proportional to $C$ :
\begin{equation}
C\ =\ \frac{K}{S}\ \sim\ K\ ;\ \ \ \ \ S\rightarrow\ {\rm const.}
\end{equation}
In such a way, and intuitively clear, 
the connectivity $C$ increases during the development of
relations. We will show that firstly the number of small components
increases and then decreases with increasing $C$ . Great components become
important. In the end all components are ``absorbed'' into one great
component.

In our case, great and strong connected components $d(s,k)$ are of
interest. This means that a great number of relations exists $k\ge s$,
so that the components are strong connected. Strong connected components are
characterized by the appearance of cycles. For $S\rightarrow\infty$ the
frequency of cycles of the length $l$ in a graph $D(S,K)$ is:
\begin{equation}
H_C^{(l)}\ =\ \frac{C^l}{l}
\end{equation}
and for semicycles with the length $l$
\begin{equation}
H_C^{l}\ =\ \frac{\left(2C\right)^l}{2l}
\end{equation}
The number of vertices of $D(S,K)$ belonging to cycles is:
\begin{equation}
Z_C\ =\ C^3\ (1-C)\ \ \ \ {\rm for}\ S\rightarrow\infty;\ \ C<1/2
\end{equation}
and of those belonging to semicycles
\begin{equation}
Z_0\ =\ 4\ C^3\ (1-2C)\ \ \ \ {\rm for}\ S\rightarrow\infty;\ \ C<1/2.
\end{equation}

The number of vertices of $D(S,K)$ belonging to one tree in
$D(S,K)$ referred to $S$ is equal to the value given by 
Erd\"os and R\'enyi \cite{erdo1960}:
\begin{equation}
y\ =\ \lim_{S\rightarrow\infty}\ \frac{R_{K,S}}{S}\ =\
\left\{
\begin{array}{l@{\hspace{1cm}}l}
1&C\le 1/2\\
x(C)/2C&C>1/2
\end{array}
\right.
\end{equation}
with
\begin{eqnarray}
R_{K,S}&=&\sum_{s=1}^S\ s\ H_s^{s-1}\\[0.3cm]
x\ {\rm e}^{\displaystyle -x}&=&2\ C\ {\rm e}^{\displaystyle -2C}\\[0.3cm]
x(C)&=&\sum_{s=1}^\infty\ \frac{s^{s-1}}{s!}\ 
\left(2\ C\ {\rm e}^{\displaystyle -2C}\right)^s
\end{eqnarray}

\begin{figure}
\vspace{0.cm}
\centering
\includegraphics[width=1.\textwidth,angle=0]{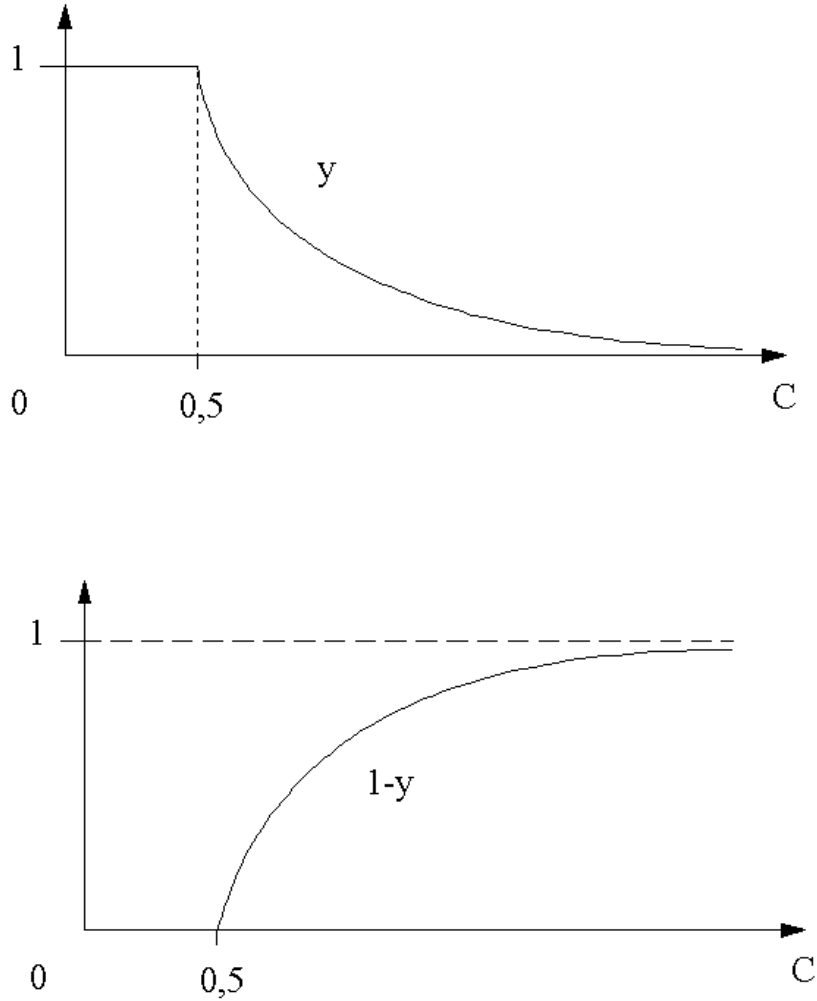}
\caption[]{The relative number of vertices of $D(S,K)$ belonging to a
tree ($y$) in dependence on the connectivity \cite{erdo1960}}
\label{fig6}
\end{figure}

The graphical representation shows analogies with a phase transition of
2$^{\rm nd}$ order -- a fact that becomes particularly obvious 
if such networks are considered by means of the percolation theory 
\cite{sonn1984app}.
For $C<1/2$ the probability is zero, that from a vertex a cycle goes out.
For $C\ge 1/2$ the probability increases up to the value one.
Opposite for $C<1/2$ a vertex belongs with the probability one
to a tree. For $C\ge 1/2$ that probability goes to zero.

For $C>1/2$ the graphs $D(S,K)$ consist of one
great component. For $C>1/2$ with growing $C$ value, one
tree after the other is ``absorbed'' by the great component. With an
increasing number of arcs ever more trees are linked
up with one great component until, finally, all vertices are linked with each
other (Fig.\ \ref{fig6}) .

\begin{figure}
\vspace{0.cm}
\centering
\includegraphics[width=.9\textwidth,angle=0]{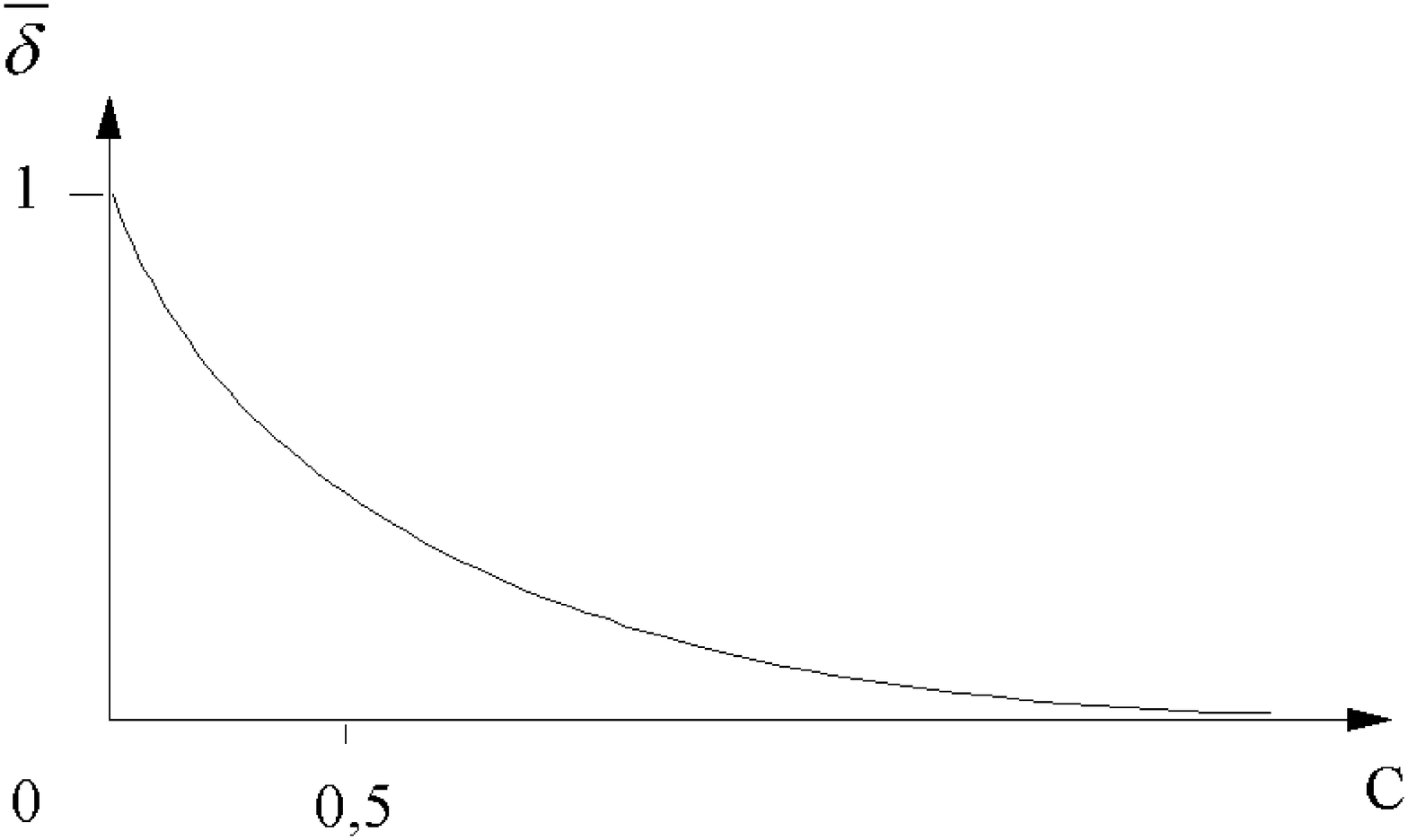}
\caption[]{The relative number of components $\bar\delta$ of a graph in
dependence on $C$}
\label{fig7}
\end{figure}

The number of components in directed graphs also corresponds to the value
given by Erd\"os and R\'enyi \cite{erdo1960} 
for undirected graphs (Fig.\ \ref{fig7}):
\begin{equation}
\bar\delta\ =\ \lim_{\frac{K}{S}\rightarrow C}\ \frac{\delta}{C}\ =\
\left\{
\begin{array}{l@{\hspace{1cm}}l}
1-C&0\le C\le 1/2\\
\frac{1}{2C}\left(x(C)-\frac{x^2(C)}{2}\right)&C>1/2
\end{array}
\right.
\end{equation}

For the examination of the development of socio-economic connections, it is
interesting to inspect the share of the graph, that is structured. If we consider
a network build from firms the number of the isolated vertices (firms) 
can be determined. In particular, they are those 
which do not belong to the connection net. In the same way,
the number of components (without isolated vertices) 
can be obtained. This is the part of 
the graph that includes the economic or knowledge exchange network.

If we understand economic connections as a manifold of possibilities, we
can arrive at some conclusions. In particular, we will observe the network of
interaction between firms over time. We keep the number of firms
(vertices) in the network constant and let the connections between them develop, i.e.
$K$, the number of edges, will increase.
On the other side $K$ is proportional to $C$. In accordance to intuition, the
connectivity rises if connections develop. As the calculation shows,
at first the number of the little components rises and then it decreases with
the further rising of $C$. 
The number of the great components rises up to the moment
when all are absorbed in one. Inside this great component the degree of
connectedness of the network plays a role. A sign for this strong
connectedness is the occurrence of cycle in the graph. As we have shown,
the frequency of cycles of the length $l$ is a simple function of the
connectivity $C$. Likewise, the probability can be given that a cycle
comes from a vertex. For $C< 1/2$ the probability is zero, that means that
the number of the vertices is unimportant. By a rising development of
connections ($C> 1/2$) this probability rises up to 1.

These findings have consequences for the development of information flows in
growing socio-economic networks. The question is if a certain type of
developing connections can also be observed in empirical data about
information flows. For instance, inside or between organizations of firms.
One possible method to trace information inside of organizations is the use
of e-mail traffic. In a recent paper 
Tyler, Wilkinson and Huberman \cite{tyle2003}
used e-mail data to map the communication network inside an industrial
organization. In this case the nodes represent people and the connections
between them are given by e-mail correspondence. They identified components
of different size and interpreted the smaller components as expressions of
communities in practice. As far as they reported, the structure of the e-mail
networks shows similarities to the structure of the organization. In addition, 
leadership roles can be identified this way.
To relate such empirical observations to the results repeated above,
one would need to look at the temporal evolution of components
and connectivity in time.

Another recent approach using percolation theory to describe socio-eco\-no\-mic
change was proposed by Silverberg and co-authors 
\cite{silv2002perc,silv2002discr}. 
Starting with a space of discrete technologies which
have a certain performance, they looked for spanning connected paths in this
technological landscape. In this picture, invention can be visualized as
isolated islands located ahead of a technological frontier which moves
forward. Under the critical value there will only be finite connected sets
(clusters or components) of different technologies, and technological change
might come to an end. If we remember that firms are the carriers for
different technologies, there is a direct link to the concept of a network of
firms exchanging information about different technologies and doing search
processes in such an abstract technological space. Then, the different
components would be related to groups of firms sharing certain technological
knowledge or being part of one chain of production. The increase in
connectivity characterizes the diffusion of a certain technology in a
specific industrial sector or across sectors.

If we consider a firm, and do not look at the whole graph, additional 
aspects can
be included: If we understand the concept of the economic connections as a
sum of abilities which the firm has acquired, we can write in simple terms:

\vspace{0.3cm}

\hspace{1cm}competence $=\ \sum$ of the abilities

\vspace{0.3cm}

\noindent
If a firm or plant creates a new economic connection, 
it can be in interaction with
others. In the picture of the graph it will be represented, as mentioned
previously, through a chain from $i$ to $j$.

\begin{figure}[h]
\centering
\unitlength1.mm
\begin{picture}(60,18)
\put(20,10){\circle*{1.5}}
\put(20,5){{\large $i$}}
\put(40,10){\circle*{1.5}}
\put(40,5){{\large $j$}}
\put(20,10){\vector(1,0){19.25}}
\put(28,14){{\large $a_{ij}$}
}
\end{picture}
\caption[]{Edge, which goes from $i$ to $j$}
\label{fig8}
\end{figure}
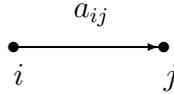

\noindent
The firm $i$ builds a connection to firm $j$. 
The connection from $i$ to $j$ is
described by $a_{ij}$. The elements $a_{ij}$ build the adjacency
matrix. If the
sum of economic connections from $i=\sum_j a_{ij}$
can be represented in a specific way we speak of firm $i$ as
 a source in the network (Fig.\ \ref{fig9}). The sum of connections 
corresponds to the sum of
arcs which leave the vertex $i$. This is exactly the sum
over the elements of row $i$ of the adjacency matrix $A$.

\begin{figure}[h]
\centering
\unitlength1.mm
\begin{picture}(37,25)
\put(20,10){\circle*{1.5}}
\put(22,6){{\large $i$}}
\put(10,10){\circle*{1.5}}
\put(5,9){{\large $m$}}
\put(20,20){\circle*{1.5}}
\put(19.5,22){{\large $k$}}
\put(27.07,17.07){\circle*{1.5}}
\put(29,18){{\large $j$}}
\put(12.93,17.07){\circle*{1.5}}
\put(10,18){{\large $l$}}
\put(12.93,2.93){\circle*{1.5}}
\put(9,0){{\large $n$}}
\put(20,10){\vector(-1,0){9.25}}
\put(20,10){\vector(0,1){9.25}}
\put(20,10){\vector(1,1){6.82}}
\put(20,10){\vector(-1,1){6.82}}
\put(20,10){\vector(-1,-1){6.82}}
\end{picture}
\vspace{0.2cm}
\caption[]{Source}
\label{fig9}
\end{figure}

\noindent
Connections which related a set of other firms
to the firm i can also be illustrated in
the graph-picture (Fig.\ \ref{fig10}).
The sum of these connections is equal $\sum_j a_{ji}$. This corresponds to the
sum of the elements of the column $i$ of the adjacency matrix.

\begin{figure}[h]
\centering
\unitlength1.mm
\begin{picture}(37,25)
\put(20,10){\circle*{1.5}}
\put(22,6){{\large $i$}}
\put(10,10){\circle*{1.5}}
\put(5,9){{\large $m$}}
\put(20,20){\circle*{1.5}}
\put(19.5,22){{\large $k$}}
\put(27.07,17.07){\circle*{1.5}}
\put(29,18){{\large $j$}}
\put(12.93,17.07){\circle*{1.5}}
\put(10,18){{\large $l$}}
\put(12.93,2.93){\circle*{1.5}}
\put(9,0){{\large $n$}}
\put(10,10){\vector(1,0){9.25}}
\put(20,20){\vector(0,-1){9.25}}
\put(27.07,17.07){\vector(-1,-1){6.82}}
\put(12.93,17.07){\vector(1,-1){6.82}}
\put(12.93,2.93){\vector(1,1){6.82}}
\end{picture}
\vspace{0.2cm}
\caption[]{Sink}
\label{fig10}
\end{figure}

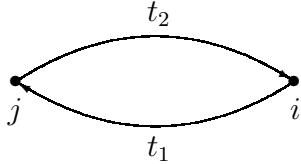
\begin{figure}[h]
\centering
\unitlength1.mm
\begin{picture}(40,30)
\put(1.5,15){\circle*{1.5}}
\put(38.5,15){\circle*{1.5}}
\linethickness{0.15mm}
\bezier{500}(1.5,15)(20,27)(38.5,15)
\bezier{500}(1.5,15)(20,3)(38.5,15)
\put(35,17){\vector(2,-1){3}}
\put(5,13){\vector(-2,1){3}}
\put(0.5,10){\large $j$}
\put(38,10){\large $i$}
\put(19,5){\large $t_1$}
\put(19,23){\large $t_2$}
\end{picture}
\caption{Mutual connections}
\label{fig11}
\end{figure}

\noindent
Connections which are based on mutual relations at different times, can be
described in the graph-picture by establishing two arcs between two vertices
(Fig.\ \ref{fig11}).

\noindent
In that picture the sum of these connections is proportional to the
sum of the cycles which leave the vertex $i$. So we can get
the number of cycles of the length $l$ from the  $l$th power of the
adjacency matrix $A^l$.

Finally let us underline that this is just a very small part of the
network theory which we explained here. We concentrated on topics relevant
to our task of describing selforganization and evolution processes.

\section{Innovation Processes and Other Stochastic Effects in Networks}

\subsection{Overview of Stochastic Effects in Networks}

\subsubsection{\label{sec3.1.1}Birth and Death Processes}

A special stochastic process called ``birth and death process'' is of
particular importance to our study of stochastic effects in economic
processes, and in particular in innovation processes. A birth process
is a random appearance of a new element in a system. A death
process is the disappearance of an element. Processes of this kind play a
big role in biology, ecology and sociology.

Processes of this type are also relevant to the field
of economic processes \cite{kani2000}.
Economic growth is characterized by structural
changes based on the introduction of new technologies in the economic world.
To describe technological evolution, one has to determine the
system, the elements and their interactions. Here we consider plants and
technologies. We consider firms composed from different plants.
The plants are introduced as elementary units which play the
role of decision carriers according to market conditions (choosing a new
technology or not). Plants also play the role of users of a particular
technology. The technologies are understood as the different types present 
in the system. The plants are the elements or representatives of these
technologies. In this perspective, eventually technologies are competing for
plants using them. This perspective differs from the way one usually thinks
about technological change, where the firm is central. The underlying
process is still a decision made by plants or firms. 
However, the model approach constructs
an inverse perspective on it. Let us note here that this perspective is
quite normal for any population dynamic approach which deals with types
(groups or species) and elements (individuals). However, in contrast to
biological processes, human beings, organizations and firms are not bound to
a certain type or group they first belong to. In contrast to individuals of
biological species they have the opportunity to change the group they belong
to. It is this kind of transition behaviour that makes the model 
particularly relevant for socio-economic applications. Let us note further
that we deliberately use 
the notion of a plant or production unit as the simplest element in
the system. By assuming that firms consist of several plants or production
units, growth processes of firms are also covered by the model approach.
Technological change is usually considered as a macroeconomic change
process. However, in order to describe it as an evolutionary process, one has
to consider this process at the microscopic level. This means we have to
consider the microeconomic carriers of technological changes. In the
framework we present here, these are the plants 
\cite{bruc1996b}. 

The basic ideas for the modelling of these processes go back to
so-called urn-models. Already in 1907 the physicists
Paul and Tatyana Ehrenfest developed a simple model for the diffusion of $N$
molecules \cite{ehre1907}. The Ehrenfests studied two urns, A and 
B, which were isolated with
respect to exchange with their surroundings. With respect to exchange 
between urns the
Ehrenfests assumed permeable connections between A and B. Because of the
isolation of the two urns, the total number of molecules in A and B remains
constant. At regular time-intervals, a molecule is randomly (that means
with the probability ($1/N$)) chosen and changes from its urn to the other
urn. 

In 1926 Kohlrausch and Schr\"odinger \cite{kohl1926} gave a continuous
diffusion-approximation for such processes. Feller \cite{fell1951}
formulated a
realistic variant of this model. He used a discrete Markovian process with
continuous time. The time between the molecule crossings was exponentially
distributed.

Originally developed for molecular processes, the model soon found many
applications to biological processes. Surveys of biological applications of
birth and death type processes were given by Bartholomay 
\cite{bart1958a,bart1958b,bart1959} 
and Eigen \cite{eige1971}.

The model of Ehrenfest represents the prototype for the investigation of
decision processes in a group between the possibilities A and B respectively
between yes and no \cite{ebel2000physica}. For example, we
may consider the decision to accept a new technology or not. Applications to
social and economic processes were surveyed by Weidlich 
\cite{weid2000}.

\subsubsection{Stochastic Effects in Small and Sensitive Networks}

Many complex systems display a surprising degree of tolerance to errors. The
results indicate a strong correlation between robustness and network
topology. In particular, scale free networks are more robust than random
networks against random node failures, but are more vulnerable when the most
connected nodes are targeted 
\cite{albe2000nat}. 

In small networks any nodes or edges play a specific role. Their addition
or removal drastically changes the properties of the whole system. Another
problem where stochastic effects play a big role is the question of how a
single new mutant can win the selection process. If one considers networks
of web sites and competition processes between them about attracting visitors
one would ask how can a new web site become a
giant cluster among other already important web sites (clusters)? Is it
possible to overcome the ``once-forever'' selection
behaviour by stochastic effects?

The addition or removal of sensitive nodes or edges is a subject of
investigation in big networks. With sensitive we mean here elements which
play a special role in the network. Let us refer to some examples.

Cellular networks can be subject to random errors as a result of mutations
or protein misfolding, as well as harsh external conditions eliminating
essential metabolites. Jeong et al.\ \cite{jeon2000}
studied the responses of the
metabolic networks of several organisms to random and preferential node
removal. Removing up to 8\% of the substrates, they found that the average
path length did not increase when nodes were removed randomly. 
However, it increases rapidly after the removal of the most-connected 
nodes and up to 500\% when only 8\% of the nodes are removed. Similar
results have been obtained for the protein network of yeast as well 
\cite{jeon2001,voge2000}. 

Sol\'e and Montoya \cite{sole2001}
studied the response of the food webs to the
removal of species (nodes) (see also \cite{mont2002}).
The results indicate that random species removal causes  
the fraction of species
contained in the largest cluster to decrease linearly. However, when the
most connected (keystone) species are successively removed, the relative
size of the largest cluster quickly decays.

The error and attack tolerance of the Internet and the World Wide Web was
investigated by Albert, Jeong and Barab\'asi 
\cite{albe2000nat}. The Internet is
occasionally subject to hacker attacks targeting some of the most
connected nodes. They show that the average path length on the internet is
unaffected by the random removal of as many as 60\% of the nodes, while if
the most connected nodes are eliminated (attack), the average path length
peaks at a very small fraction of removed nodes. Albert, Jeong and
Barab\'asi investigated the World Wide Web \cite{albe1999}
and showed that the
network survives as a large cluster under high rates of failure, but under
attack, the system abruptly crashes. These authors write: ``The result is
that scale-free networks display a high degree of robustness against random
errors, coupled with a susceptibility to attacks.''

Wagner and Fell \cite{wagn2000}
studied the clustering coefficient, focusing on the
energy and biosynthesis metabolism of the Escherichia coli bacterium. They
found that in addition to the power-law degree distribution, the undirected
version of this substrate graph has a small average path length and a large
clustering coefficient.

Bianconi and Barab\'asi \cite{bian2001prl} 
showed the existence of a closed link
between evolving networks and an equilibrium Bose gas. According to them, the
mapping to a Bose gas predicts the existence of two distinct phases as a
function of the energy distribution. In the fit-get-rich-phase, the fitter
nodes acquire edges at a higher rate than older but less fit nodes. In the
end, the fittest node will have the most edges, but the richest node is not
an absolute winner, since its share of the edges decays to zero for large
system size.

Maurer and Huberman \cite{maur2000}
present a dynamic model of web site growth in
order to explore the effects of competition among web sites. They show that
under general conditions, as the competition between sites increases, the
model exhibits a sudden transition from a regime in which many sites thrive
simultaneously, to a ``winner takes all market'' in which a few sites grab
almost all the users, while most other sites become nearly extinct. This
prediction is in agreement with empirical data measurements on the nature
of electronic markets.

Dorogovtsev, Mendes and Samukhin \cite{doro2002}
developed a statistical mechanic
approach for random networks. They summarize: ``Using the traditional
formalism of statistical mechanics, we have constructed a set of equilibrium
statistical ensembles of random networks without correlation and have found
their partition function and main characteristics. We have shown that a
``scale-free'' state in equilibrium networks without condensate may exist
only in a single marginal point, so that in such an event this state is an
exhibition.'' They underline the important fact that this differs crucially
 from the situation for growing networks. The latter, while growing,
self-organize into scale-free structures in a wide range of parameters
without condensation.

\subsection{Stochastic Analysis of Innovation Processes}

\subsubsection{\label{sec3.2.1}A General Formulation of the Model}

The stochastic approach given here is based on a model, which was developed
in the context of general models of evolutionary processes, and in particular
biochemical processes \cite{ebel1981,schi1981,ebel1982,ebel1986bio}.
Later, the model found numerous applications in modelling scientific
evolution \cite{ebel1986czech,bruc1990} 
and technological evolution \cite{bruc1994,bruc1996b}.
In this paper, we first introduce the model framework in a general form
and later concentrate on its application to economic innovation and
competition processes.

In the stochastic picture we use the ideas developed in subsection 
\ref{sec1.1} and in the subsection \ref{sec3.1.1}
about the occupation number space. Opposite to the
deterministic models, the stochastic description offers the advantage that at
finite times new types (sorts, fields, species, technologies) can arise or
``die out''.

Let us introduce a set of types numbered by $i=1,2,\ldots ,s$. We denote by
$N_i (t)$ the number of elements belonging to a certain type. For an economic
application $N_i (t)$ represents the number of plants using the technology
$i$ (Note, that a type corresponds to an urn in the 
Ehrenfest problem formulation.).
These numbers are called occupation numbers. They are a function of time.
The occupation numbers are positive or zero.
\begin{equation}
N_i(t)\  =\ \{0,1,2,\cdots\}\ .
\end{equation}
Now, the state of the system at the time $t$ can be described 
by the probability distribution of the occupation numbers
\begin{equation}
P(N_1,N_2,\ldots,N_s;t)\ =\ P(N;t)
\end{equation}
We consider as elementary processes -- processes during which only one
occupation number can change and as transition processes -- processes during
which at most two occupation numbers can change:
\begin{equation}
\begin{array}{c@{\ \ \ \longrightarrow\ \ \ }c}
(N_i)&(N_i + 1)\\[0.3cm]
(N_i)&(N_i - 1) \\[0.3cm]
 \left( 
\begin{array}{c}
 N_i\\[0.3cm]
 N_j
\end{array}\right)&
 \left( 
\begin{array}{c}
 N_i - 1\\[0.3cm]
 N_j + 1
\end{array}\right)
\end{array}
\end{equation}
For the time being, we also assume that growth and decline processes of the
total number of elements in the system are possible. However, let us note
here that the original formulation of the Ehrenfest model only contains
transition processes.  This is due to the fact 
that in the Ehrenfest model the total number of elements remains 
constant and only exchange
between the urns occurs. Decision processes in the model occur as
transitions of elements between types. 

If we assume that all decisions
leading to a change of the set of the occupation numbers depend mostly on
the present state, we can apply the concept of Markov process. 
Then, we can describe the dynamics
of the system with the help of the master equation. This
equation is a balance equation between building and reduction processes:
\begin{equation}
\frac{\partial P(N;t)}{\partial t}\ =\ W(N\vert N^\prime)\
P(N^\prime)\ -\ W(N^\prime\vert N)\ P(N)
\end{equation}
with
\begin{equation}
N\ =\ \left\{ N_1,N_2,\ldots,N_s\right\}\ .
\end{equation}
The transition probabilities per time unit that the system turns from the
state $N^\prime$ to the state $N$ or vice versa are expressed by
$W(N\vert N^\prime)$ respectively $ W(N^\prime\vert N)$.

The transition probabilities are supposed as follows 
\cite{feis1977,ebel1982,jime1980,ebel1986bio,hein1981}.

\noindent
1.\ Spontaneous generation
\begin{eqnarray}
&&\hspace{-1cm}W(\ldots,N_i + 1,\ldots,N_j,\ldots,N_k,\ldots\vert
\ldots,N_i,\ldots,N_j,\ldots,N_k,\ldots)\nonumber\\[0.3cm]
&=&A_i^{(0)}
\label{eq24}
\end{eqnarray}
2.\ Self-reproduction
\begin{eqnarray}
&&\hspace{-1cm}W(\ldots,N_i + 1,\ldots,N_j,\ldots,N_k,\ldots\vert
\ldots,N_i,\ldots,N_j,\ldots,N_k,\ldots)\nonumber\\[0.3cm]
&=&A_{ij}^{(1)}\ N_j\ +\ E_i^{(1)}\ N_i\\[0.3cm]
E_i^{(1)}&=& A_i^{(1)}\ +\ B_{ij}^{(1)}\ N_j\ +\  C_{ijk}^{(1)}\ N_j N_k
\label{eq2526}
\end{eqnarray}
3.\ Decay
\begin{eqnarray}
&&\hspace{-1cm}W(\ldots,N_i - 1,\ldots,N_j,\ldots,N_k,\ldots\vert
\ldots,N_i,\ldots,N_j,\ldots,N_k,\ldots)\nonumber\\[0.3cm]
&=&E_i^{(2)}\ N_i\\[0.3cm]
E_i^{(2)}&=& A_i^{(2)}\ +\ B_{ij}^{(2)}\ N_j
\label{eq2728}
\end{eqnarray}
4.\ Conversion/Transition/Exchange/Mutation
\begin{eqnarray}
&&\hspace{-1cm}W(\ldots,N_i + 1,\ldots,N_j - 1,\ldots,N_k,\ldots\vert
\ldots,N_i,\ldots,N_j,\ldots,N_k,\ldots)\nonumber\\[0.3cm]
&=&E_j^{(3)}\ N_j\\[0.3cm]
E_j^{(3)}&=&A_{ij}^{(3)}\ +\ B_{ij}^{(3)}\ N_i
\ +\bar{B}_{ik}^{(3)} N_k \ +\  C_{ijk}^{(3)}\ N_i N_k
\label{eq2930}
\end{eqnarray}
with $j\ne i; k\ne i;j$.

The coefficients can be differently introduced for special cases.
For instance, they can considered as constant or as being functions of the
total number of elements and other the system parameters. Later approach
introduces an additional non-linearity to the system.

The content of the four elementary processes introduced above will be quite
different according to the nature of the system under consideration. For
instance, for catalytic networks, self-reproduction may appear as result of a
process of spontaneous self-reproduction (term related to $A_i^{(1)}$),
error reproduction ($B_{ij}^{(1)}$) or catalytic self-reproduction 
($C_{ijk}^{(1)}$). The
decay can appear in the form of spontaneous decay ($A_i^{(2)}$) and of decay 
related to catalytic help ($B_{ij}^{(2)}$). 
Transition or conversion processes in catalytic
networks correspond to mutation processes with reproduction and ternary
reproduction processes, which play a role with regard to processes with
constant overall particle number.

In the case of technological evolution, self-reproduction appears as a growth
process of firms expanding their number of production units and plants using
the same technology. Synergetic effects from the surrounding network of
firms are supposed to take place when the transition rate also depends on
the number of firms using another technology. Spontaneous generation stands
for startups. Decay processes both refer to a decrease of firm size (closing
of production units) and a closing down of firms. The most interesting
process is related to conversion or transition. Here, the use of another
technology by a production unit is described. The use of a technology
may be new for the firm only (firm-specific innovation) or for the whole
system of firms (system-specific innovation). Triggered by R\&D, both
invention of a technology and imitation behaviour is covered by this
process. Also in this case, other technologies might influence the decision
of a firm for a certain technology. The advantage of this type of model is
that technological change is considered as the outcome of the development of
a network of technologies and firms influencing each other.

The transition probabilities are formulated generally. As a special case we
can get from this {\it Ansatz} the stochastic equations, which correspond to the
deterministic {\sc Eigen}-model \cite{eige1978c} 
with the condition of constant overall particle number 
\cite{jime1980,ebel1981,hein1981}.

With the help of the $s$-dimensional generation function
\begin{equation}
F(s_1,s_2,\ldots,s_s;t)\ =\ \sum_N\ s_1^{N_1}\ s_2^{N_2}\ \ldots\ 
s_s^{N_s}\ P(N_i,t)\ \ \ {\rm with}\ \ \vert s_i\vert\ <\ 1
\end{equation}
we can write the master equation with the transition probabilities as
follows:
\begin{eqnarray}
\dot{F}(s;t)&=&\sum_{i\ne j}\left\{ A_i^{(0)}\ (s_i-1)\ F\ +\ 
 A_i^{(1)}\ s_i (s_s-1)\ \frac{\partial F}{\partial s_i}\right.
\nonumber\\[0.3cm]
&&\ +\ A_{ij}^{(1)}\ s_j (s_i-1)\ \frac{\partial F}{\partial s_j}\ +\
B_{ij}^{(1)}\ s_i s_j (s_i-1)\ \frac{\partial^2 F}{\partial s_i\partial s_j}
\nonumber\\[0.3cm]
&&\ +\ C_{ijk}^{(1)}\ s_i s_j s_k (s_i-1)\ 
\frac{\partial^3 F}{\partial s_i\partial s_j\partial s_k}\ +\
A_i^{(2)}\ (1-s_i)\ \frac{\partial F}{\partial s_i}\nonumber\\[0.3cm]
&&\ +\ 
B_{ij}^{(2)}\ s_j (1-s_i)\ \frac{\partial^2 F}{\partial s_i\partial s_j}
\ +\ A_{ij}^{(3)}\ (s_i-s_j)\ \frac{\partial F}{\partial s_j}\nonumber\\[0.3cm]
&&\ +\ 
B_{ij}^{(3)}\ s_i (s_i-s_j)\ \frac{\partial^2 F}{\partial s_i\partial s_j}
\ +\ 
\bar{B}_{ik}^{(3)}\ s_k (s_i-s_j)\ 
\frac{\partial^2 F}{\partial s_k\partial s_j}\nonumber\\[0.3cm]
&&\left.\ +\ C_{ijk}^{(3)}\ s_i s_k (s_i-s_j)\ 
\frac{\partial^3 F}{\partial s_i\partial s_j\partial s_k}\right\}
\end{eqnarray}

\subsubsection{A Network Representation of the Model}

In order to relate the model to the idea of sensitive
networks we now introduce a network representation of 
transition probabilities given in (\ref{eq24})-(\ref{eq2930}). 
We consider a system with $s$ interacting types (sorts, fields,
plants). This system can be described by a graph in which
each element $i$ corresponds to a vertex of the number $i$. We mark the
transition probabilities for different processes by edges of different type
in order to distinguish these probabilities:
\begin{eqnarray}
\begin{array}{ll@{\hspace{1cm}}l@{\hspace{0.5cm}}l}
1.&\parbox[t]{5cm}{Spontaneous generation (simple innovation)}&
A_i^{(0)}&
\unitlength1.mm
\begin{picture}(20,5)
\put(5,1){\circle*{1.5}}
\put(5,-3){\it i}
\put(15,1){\vector(-1,0){9.25}}
\end{picture}
\\[0.8cm]
2.&\parbox[t]{5cm}{Self-reproduction}&
A_i^{(1)} N_i&
\unitlength1.mm
\begin{picture}(20,5)
\put(10,1){\circle*{1.5}}
\put(10,-3){\it i}
\put(10,4){\circle{6}}
\put(8,1.8){\vector(4,-1){1.5}}
\end{picture}
\\[0.8cm]
&\parbox[t]{5cm}{Error reproduction}&
A_{ij}^{(1)} N_j&
\unitlength1.mm
\begin{picture}(20,5)
\put(5,1){\circle*{1.5}}
\put(5,-3){\it j}
\put(15,1){\circle*{1.5}}
\put(15,-3){\it i}
\put(5,1){\line(1,0){1.5}}
\put(7,1){\line(1,0){1.5}}
\put(9,1){\line(1,0){1.5}}
\put(11,1){\line(1,0){1.5}}
\put(13,1){\vector(1,0){1.25}}
\end{picture}
\\[0.8cm]
&\parbox[t]{5cm}{Catalytic self-reproduction \hfill

(sponsored self-reproduction)}&
\left\{ 
\begin{array}{l}
B_{ij}^{(1)} N_i N_j\\[0.8cm]
C_{ijk}^{(1)} N_i N_j N_k
\end{array}\right.
&
\unitlength1.mm
\begin{picture}(20,5)
\put(5,7){\circle*{1.5}}
\put(5,3){\it i}
\put(15,7){\circle*{1.5}}
\put(15,3){\it j}
\put(5,7){\line(1,0){1.5}}
\put(7,7){\line(1,0){1.5}}
\put(9,7){\line(1,0){1.5}}
\put(11,7){\line(1,0){1.5}}
\put(13,7){\line(1,0){1.5}}
\put(5,10){\circle{6}}
\put(3,7.8){\vector(4,-1){1.5}}
\put(5,-6){\circle*{1.5}}
\put(5,-10){\it i}
\put(15,-6){\circle*{1.5}}
\put(18,-7){\it j}
\put(5,-6){\line(1,0){1.5}}
\put(7,-6){\line(1,0){1.5}}
\put(9,-6){\line(1,0){1.5}}
\put(11,-6){\line(1,0){1.5}}
\put(13,-6){\line(1,0){1.5}}
\put(5,-3){\circle{6}}
\put(3,-5.2){\vector(4,-1){1.5}}
\put(15,0){\circle*{1.5}}
\put(18,-1){\it k}
\put(5,0){\line(1,0){1.5}}
\put(7,0){\line(1,0){1.5}}
\put(9,0){\line(1,0){1.5}}
\put(11,0){\line(1,0){1.5}}
\put(13,0){\line(1,0){1.5}}
\end{picture}
\\[1.2cm]
3.&\parbox[t]{5cm}{Spontaneous decay}&
A_i^{(2)} N_i&
\unitlength1.mm
\begin{picture}(20,5)
\put(5,1){\circle*{1.5}}
\put(5,-3){\it i}
\put(5,1){\vector(1,0){10}}
\end{picture}
\\[0.8cm]
&\parbox[t]{5cm}{Catalytic decay}&
B_{ij}^{(2)} N_i N_j&
\unitlength1.mm
\begin{picture}(20,5)
\put(15,1){\circle*{1.5}}
\put(15,-3){\it j}
\put(5,6){\circle*{1.5}}
\put(5,7){\it i}
\put(5,6){\vector(0,-1){10}}
\put(5,1){\line(1,0){1.5}}
\put(7,1){\line(1,0){1.5}}
\put(9,1){\line(1,0){1.5}}
\put(11,1){\line(1,0){1.5}}
\put(13,1){\line(1,0){1.5}}
\end{picture}
\\[0.8cm]
4.&\parbox[t]{5cm}{Mutation (innovation)}&
A_{ij}^{(3)} N_j&
\unitlength1.mm
\begin{picture}(20,5)
\put(5,1){\circle*{1.5}}
\put(5,-3){\it j}
\put(15,1){\circle*{1.5}}
\put(15,-3){\it i}
\put(5,1){\vector(1,0){9.25}}
\end{picture}
\\[0.8cm]
&\parbox[t]{5cm}{Mutation (innovation) with reproduction}&
\left\{ 
\begin{array}{l}
B_{ij}^{(3)} N_i N_j\\[0.8cm]
C_{ijk}^{(3)} N_i N_j N_k
\end{array}\right.
&
\unitlength1.mm
\begin{picture}(20,5)
\put(5,7){\circle*{1.5}}
\put(5,3){\it i}
\put(15,13){\circle*{1.5}}
\put(18,12){\it j}
\put(15,13){\vector(-1,0){10}}
\put(5,10){\circle{6}}
\put(3,7.8){\vector(4,-1){1.5}}
\put(5,-6){\circle*{1.5}}
\put(5,-10){\it i}
\put(15,-6){\circle*{1.5}}
\put(18,-7){\it k}
\put(5,-6){\line(1,0){1.5}}
\put(7,-6){\line(1,0){1.5}}
\put(9,-6){\line(1,0){1.5}}
\put(11,-6){\line(1,0){1.5}}
\put(13,-6){\line(1,0){1.5}}
\put(5,-3){\circle{6}}
\put(3,-5.2){\vector(4,-1){1.5}}
\put(15,0){\circle*{1.5}}
\put(18,-1){\it j}
\put(15,0){\vector(-1,0){10}}
\end{picture}
\\[1.2cm]
&\parbox[t]{5cm}{Mutation without reproduction}&
\bar{B}_{ik}^{(3)} N_k N_j&
\unitlength1.mm
\begin{picture}(20,5)
\put(5,-6){\circle*{1.5}}
\put(5,-10){\it k}
\put(15,-6){\circle*{1.5}}
\put(15,-10){\it i}
\put(15,4){\circle*{1.5}}
\put(15,6){\it j}
\put(5,-6){\line(1,0){1.5}}
\put(7,-6){\line(1,0){1.5}}
\put(9,-6){\line(1,0){1.5}}
\put(11,-6){\line(1,0){1.5}}
\put(13,-6){\vector(1,0){1.25}}
\put(15,4){\vector(0,-1){9.25}}
\end{picture}
\end{array}\nonumber\\[0.5cm]
&&\nonumber
\end{eqnarray}
For under-occupied systems the transition from a non-occupied ($N_i=0$) 
state to an occupied ($N_i>0$) state is of special
interest. We will call such a transition an innovation. To describe this in
the network picture we distinguish two states for one vertex. The states are
marked as follows:
\unitlength1.mm
\begin{picture}(120,25)
\put(20,15){\circle{2}}
\put(22,12){i}
\put(40,14){\rm non-occupied}
\put(80,14){$N_i\ =\ 0$}
\put(20,5){\circle*{2}}
\put(22,2){i}
\put(40,4){\rm occupied}
\put(80,4){$N_i\ >\ 0$}
\end{picture}
We also distinguish two states for the edges of the graph. Edges which go
out from unoccupied vertices ($N_i=0$) are omitted, because
they are inactive (they cannot work). Active (working) edges are
characterized by the graphic representation shown above. For socio-economic
networks, non-occupied vertices stay for new technologies, which have not yet
discovered. They can be understood as hidden possibilities. The model
does not allow for the prediction of a certain new technology but it can make 
statements how the system
handles the appearance of new technologies in general.

Eventually, we get a graph which describes the whole system at a fixed time
$t$. If the colour of a vertex is changed, new connections (elementary
processes) can flare up or former connections can no longer continue. Processes
can spread the elements on the non-occupied vertices or they can select them
 from the occupied vertices. The basic structure of the network is the
``maximal'' graph (all vertices $i$ are occupied, all reactions can work).

In a network with a small overall number of elements (individuals,
organizations, plants) ($N\ll s$), a lot of vertices are not
occupied. Let us consider a graph with ($N\gg s$) and reduce the
overall number of elements to a state ($N\ll s$). In general, vertices 
such as i and j become
non-occupied by the following processes:
$A_i^{(2)} N_i$, $B_{ij}^{(2)} N_i N_j$, $A_{ij}^{(3)} N_j$, $B_{ij}^{(3)} N_i N_j$,
\begin{math} \bar{B}_{ik}^{(3)} N_k N_j \end{math}, $C_{ijk}^{(3)} N_i N_j N_k$.
These are the processes of decay and the processes of
conversion. If only a certain part of the process works because of the
small overall number, we get a graph with a lot of components. With
a decreasing number of elements (individuals, organizations, plants) the
maximal graph with few components develops to a graph with a lot of
components \cite{ebel1982}. The ``minimal'' graph we can get
(maximally decomposed), consists of self-reproduction processes, spontaneous
generation (simple innovations), sponsored innovation-processes and
sponsored innovation-processes with self-reproduction (self-reproducing
process). Vice versa, vertices can be occupied by the following processes:
spontaneous generation (simple innovation) $A_i^{(0)}$,
error reproduction $A_{ij}^{(1)} N_j$, mutation without
reproduction (innovation without self-reproduction) $A_{ij}^{(3)} N_j$, 
and mutation without reproduction but with catalytic
help (sponsored innovation without self-reproduction) 
$\bar{B}_{ik}^{(3)} N_k N_j$.

We assume that the processes through which the vertices can be occupied are
very rare. This means spontaneous generations (simple innovations) 
and mutations have a
small probability. After a relatively short time the components are in a
local equilibrium (first process). Then, sponsored innovation-processes play
a role starting from an occupied component which is in equilibrium. New
components (non-occupied) can be occupied (second process). This is a
hopping process between components. Certain components die out
under selection pressure, others survive and growth.

\subsection{Application to Technological Innovations}

\subsubsection{Technologies with Linear Growth Rates}

In the following, we consider a system in which the total number of
elements is constant. As mentioned above,
a useful instrument to model such processes is the
theory of stochastic transitions between urns, established by Ehrenfest.
Compared with the general model introduced so far in the case of a closed
system, all elementary processes appear as transition processes. Further, let
us restrict ourselves to the case of technological evolution. Then, the urns stay
for different technologies. Symbolic spheres travelling between the urns
stand for plants looking for technologies.

$N_i$  is the number of plants, which uses the technology
$i$, this means they belong to the urn $i$. 
This numbers are called occupation numbers:
\begin{equation}
N_i(t)\ =\ \{0,1,2,\ldots\}\ .
\end{equation}
First, we formulate a simple model for the binary decision process:
\begin{equation}
N\ =\ N_1\ +\ N_2\ .
\end{equation}
We assume that during elementary processes the occupation number only
changes by $\pm1$. This is the so called one-step-process.
During transition processes at most two occupation numbers can change in
such a way:
\begin{equation}
{N_1\choose N_2}\ \ \longrightarrow\ \ {N_1-1\choose N_2+1}
\end{equation}
For instance, we assume that $E_1$ is the growth rate of
plants using technology 1. For a new technology 2 the growing rate is 
$E_2$. We assume:
\begin{equation}
E_2\ >\ E_1\ .
\end{equation}
In this case the technology 2 has a greater growth potential or will grow
faster than technology 1. We will assume that this is an expression for
technology 2 to be ``better'' or more suitable for plants. In the stochastic
picture the transition from urn 1 to urn 2 will simply have a higher
probability than the other way around. So, plants will more often change
their technology towards technology 2. They will replace the old technology
by a better (new) one.

We write the transition probability for this process as follows:
\begin{equation}
W^+(N_2 + 1\vert N_2)\ =\ E_2\ N_1\ \left(\frac{N_2}{N}\right)\ +\
E_{21}\ =\ W_{N_2}^+
\end{equation}
In particular, we assume, that the transition probability is proportional to
the number of plants that use the old technology:
\begin{equation}
W^+(N_2)\ \sim\ N_1
\end{equation}
Furthermore, the probability is also proportional to the relative number of
plants that use the new technology
\begin{equation}
W^+(N_2)\ \sim\ \frac{N_2}{N}
\end{equation}
We introduce an additional process of spontaneous change from technology 1
to technology 2 which is described by the coefficient $E_{21}$.
The opposite spontaneous transition appears with the coefficient $E_{12}$. 
In total the opposite transition process has the probability:
\begin{equation}
W^-(N_2 - 1\vert N_2)\ =\ E_1\ N_2\ \left(\frac{N_1}{N}\right)\ +\
E_{12}\ =\ W_{N_2}^-
\end{equation}
We can formulate the master equation for this discrete process. This
equation describes the time-behaviour of the probability distribution of the
occupation numbers.
\begin{equation}
P(N_1,N_2;t)
\end{equation}
With the mentioned transition probabilities follows:
\begin{eqnarray}
\frac{\partial}{\partial t}P(N_1,N_2;t)
&=&W_{N_2-1}^+(N_2\vert N_2 - 1)\ P(N_2 - 1;t)\nonumber\\[0.3cm]
&&\ +\ W_{N_2+1}^-(N_2\vert N_2 + 1)\ P(N_2 + 1;t)\nonumber\\[0.3cm]
&&\ -\ W_{N_2}^+(N_2 + 1\vert N_2)\ P(N_2;t)\nonumber\\[0.3cm]
&&\ -\ W_{N_2}^-(N_2-1\vert N_2)\ P(N_2;t)
\end{eqnarray}
and
\begin{eqnarray}
&&\hspace{-1.5cm}\frac{\partial}{\partial t}P(N_1,N_2;t)\nonumber\\[0.3cm]
&=&\left[E_{21}\ +\ \frac{E_2}{N}\ (N_2-1)(N_1+1)\right]\
P(N_1+1,N_2-1;t)\nonumber\\[0.3cm]
&&\ +\ \left[E_{12}\ +\ \frac{E_1}{N}\ (N_2+1)(N_1-1)\right]\
P(N_1-1,N_2+1;t)\nonumber\\[0.3cm]
&&\ -\ \left[E_{21}\ +\ E_{12}\ +\ (E_1+E_2)N_2(N-N_2)\right]\
P(N_1,N_2;t)
\end{eqnarray}
If we use the relation $N_1 + N_2 = N$ (i.e.,$N_1 = N - N_2$), 
we can write:
\begin{eqnarray}
&&\hspace{-1.5cm}\frac{\partial}{\partial t}P(N_2;t)\nonumber\\[0.3cm]
&=&\left[E_{21}\ +\ \frac{E_2}{N}\ (N_2-1)(N-N_2+1)\right]\
P(N_2-1;t)\nonumber\\[0.3cm]
&&\ +\ \left[E_{12}\ +\ \frac{E_1}{N}\ (N_2+1)(N-N_2-1)\right]\
P(N_2+1;t)\nonumber\\[0.3cm]
&&\ -\ \left[E_{21}\ +\ E_{12}\ +\ \frac{(E_1+E_2)}{N}N_2(N-N_2)\right]\
P(N_2;t)
\end{eqnarray}
To obtain statements for the deterministic case, we define
the mean value:
\begin{equation}
\langle N_2(t)\rangle\ = \ \sum_{N_2=0}^\infty\ N_2\ P(N_2;t).
\end{equation}
By multiplying the master equation with $N_2$ and a following
summation, we obtain:
\begin{equation}
\frac{\rm d}{{\rm d} t}\langle N_2(t)\rangle\ =\ 
\frac{E_2-E_1}{N}\ \langle N_2 (N-N_2)\rangle\ +\ (E_{21}-E_{12}).
\end{equation}
Using the approximation $\langle (N_2)^2\rangle \approx \langle N_2\rangle^2$ 
and the abbreviations
\begin{equation}
x_2\ =\ \frac{\langle N_2\rangle}{N}\ ;\ \
\alpha\ =\ E_2\ -\ E_1\ ;\ \ \beta\ =\ \frac{E_{21}-E_{12}}{N}
\end{equation}
we achieve the corresponding deterministic equations:
\begin{equation}
\frac{{\rm d}x_2}{{\rm d} t}\ =\ \alpha\ x_2\ (1-x_2)\ +\ \beta
\end{equation}
Let us compare the stationary behaviour of both the stochastic and the
deterministic model. In the case of two technologies we can explicitly
derive the stationary solution of the master equation:
\begin{equation}
P^0(N_2)\ =\ \frac{W^+(N_2)\ W^+(N_2-1)\ \cdots\ W^+(N-1)}
{W^-(N_2+1)\ W^-(N_2+2)\ \cdots\ W^-(N)}\ P^0(N)
\end{equation}
In the deterministic case the stationary solution for $E_2 > E_1$ is:
\begin{equation}
x_2\ = 1,\ \ {\rm that\ is}\ \ N_2\ = N
\end{equation}
The new (better) technology will replace the old one. If a final stable
stationary solution is reached by the system all plants will use the new
technology.

For the stochastic case we get a quite different picture. In particular, the
probability distribution $P(N_2)\ne 0$ for $N_2\ne N$. That
means, in the stochastic case the old and new technology will coexist, given
that the difference between $E_1$ and $E_2$ is not too large. 
Otherwise, the old technology will only survive in a small niche.

An interesting case occurs, if the spontaneous rates are $E_{21}=E_{12}=0$. 
In this case the states $N_1=0$, $N_2=0$ are called
absorber states. This means, the system can not leave these states. For the
stationary solution of the master equation we get:
\begin{equation}
P^0(N_2)\ =\ \sigma_1\ \delta_{0 N_2}\ +\ \sigma_2\ \delta_{N N_2}\ ;
\sigma_1\ +\ \sigma_2\ =\ 1
\end{equation}
$\sigma_i$ is a real number between zero and one, From the absorber
states $N_2=0$ and $N_1=0$ ($N_2=N$) the other states can not
be reached.

 From the initial state follows, which stationary solution is occupied by the
system. After some calculations we can give for $\sigma_2$ in the
limit $N\gg 1$ and $N_2(t=0)$ the following equation:
\begin{equation}
\sigma_2\ =\ \left\{
\begin{array}{l@{\hspace{1cm}{\rm for}\ \ \ }l}
0&E_2 < E_1\\
1-\left(\frac{E_1}{E_2}\right)^{N_2(0)}&E_2 > E_1
\end{array}
\right.
\end{equation}
with $N_2(0) = N_2(t=0)$ the initial state of the system. Is $N_2(0)$ 
the number of users at time $t=0$ of the
technology 2, then $\sigma_2$ is the probability, that for 
$t\rightarrow\infty$ $N_2 = N$ users change to technology 2. $\sigma_1$ 
is the probability that for $t\rightarrow\infty$ $N_2=0$ (i.e., $N_1 = N$)
users changes to technology 2, this means the new technology
has not survived. In general, $\sigma_i$ is the survival probability of
technology $i$.

If a small number of plants $N_2(0)$ use a new technology, this
technology will disappear if its growth rate $E_2$ is smaller
than that of the old one. If the growth rate $E_2$ of the new
technology is considerable larger, the new technology will be successful
with the probability.
\begin{equation}
1\ -\ \left(\frac{E_1}{E_2}\right)^{N_2(0)}
\end{equation}
Then, with the small probability:
\begin{equation}
\left(\frac{E_1}{E_2}\right)^{N_2(0)}
\end{equation}
the old technology will still be used in the system.

In the deterministic case a new technology with higher growth rates is
always successful. This is the case of pure Darwinian selection where the
fittest and only the fittest survives at the end of the process. 

What we
observe in socio-economic system is usually a variety of technologies
\cite{savi1996}.  This can be either explained due to the action of mutation
processes and error reproduction or due to the presence of stochastic
processes. As we showed above in the stochastic situation the new better
technology will only survive with a certain probability. Empirical studies
of growth processes in ensembles of technologies might shed  light on which
growth mechanism is present in a certain system.

\subsubsection{Technologies with Quadratic Growth Rates}

In the following we consider again the case with absorber states, 
that means we neglect the terms $E_{21}$, $E_{12}$. As an extension 
to the model introduced above, instead of the linear terms
($E_1$, $E_2$) we introduce now non-linear components 
($b_1$, $b_2$, $V$). $V$ is a parameter which in biochemical applications
stays for the spatial volume of the system. In general terms it is related to the size
of the system. Remember that we still have a system with 
a constant total number of elements (plants). So all processes appear 
as transition processes. The non-linearities might stay for processes 
of catalytic self-reproduction as well as for transition or mutation 
processes influences by other types (technologies). We can write
the transition probabilities as:
\begin{eqnarray}
W^+(N_1)&=&\frac{b_1}{N V}\ N_2\ N_1^2\\[0.3cm]
W^-(N_1)&=&\frac{b_2}{N V}\ N_2^2\ N_1
\end{eqnarray}
(compare \cite{ebel1981} 
in detail for an application to biochemical systems). We
assume that the sum of the occupation numbers remains constant:
\begin{equation}
N_1\ +\ N_2\ =\ N\ =\ {\rm const.}
\end{equation}
The probability distribution of the occupation numbers is:
\begin{equation}
P(N_1,N_2;t)
\end{equation}
Analogous to the above chapter we can formulate the master equation 
as follows:
\begin{eqnarray}
\frac{\partial}{\partial t}P(N_1;t)
&=&W^+(N_1 - 1)\ P(N_1 - 1;t)
\ +\ W^-(N_1 + 1)\ P(N_1 + 1;t)\nonumber\\[0.3cm]
&&\ -\ \left[ W^+(N_1) + W^-(N_1)\right]\ P(N_1;t)
\end{eqnarray}
By multiplying the master equation with $N_k/V$ and summing
over all occupation numbers, we can get after factorization 
of the mean values the deterministic equation:
\begin{eqnarray}
&&\ \ \ \langle N_k\rangle/V\ =\ x_k\ \ \ \ {\rm with}\ \ \
x_1\ +\ x_2\ =\ \frac{N}{V}\ =\ C\ =\ {\rm const.}\nonumber\\[0.3cm]
\dot{x}_i&=&b_i\ x_i^2\ -\ \varphi\ x_i\ ;\ \ \ i\ =\ 1, 2
\end{eqnarray}
with
\begin{equation}
\varphi\ =\ \frac{b_1 x_1^2 + b_2 x_2^2}{C}
\end{equation}
Equations of the same form have been derived for so-called 
hypercyclic system to describe the evolution of macromolecules 
\cite{eige1978c}. 
Their behaviour is very well understood. As a result of the quadratic 
terms in the growth rates the phase space is split into two regions 
separated by a separatrix $S_i$:

\begin{figure}[h]
\centering
\unitlength1.mm
\begin{picture}(80,15)
\put(20,10){\circle*{1.5}}
\put(13,3){{\large $x_i = 0$}}
\put(40,10){\circle*{1.5}}
\put(38.5,3){{\large $S_i$}}
\put(60,10){\circle*{1.5}}
\put(53,3){{\large $x_i = C$}}
\put(20,10){\line(1,0){40}}
\end{picture}
\caption[]{Phase space in the deterministic case}
\label{fig12}
\end{figure}
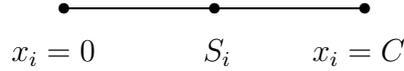

Let us also note here that because of the $N$ equal constant condition the 
dimensionality of the system reduces. Due to the nature of the transition 
process, the occupation numbers in the system change along a diagonal of
the two-dimensional phase space. In the deterministic picture, the selection 
behaviour depends on the initial conditions $x_i(t=0)$ of the system. 
A certain technology $i$ only can win  ($x_i = C$ for $t\to\infty$), 
if the initial condition places the system beyond the point of the
separatrix:
\begin{equation}
x_i(0)\ >\ S_i\ ;\ \ \ S_i\ =\ \frac{C\ b_j}{b_i + b_j}
\end{equation}
This way it is possible that a new technology $i$, even when having a 
greater growth rate, will not win because at the beginning the number 
of users (plants) is too small.
\begin{equation}
x_i(t=0)\ <\ S_i
\end{equation}
Such a situation has been called once-forever selection or   
hyperselection. In the case of macromolecular evolution this feature 
was used to explain the uniqueness of the genetic code. For 
technological evolution we have argued elsewhere that hyperselection 
is an alternative explanation for so-called lock-in phenomena
of technologies \cite{bruc1994,bruc1996b}.
The situation changes again if we look at the stochastic picture. 
Here the phase space looks as follows:

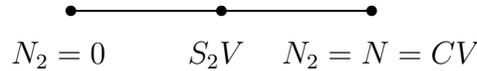
\begin{figure}[h]
\centering
\unitlength1.mm
\begin{picture}(80,15)
\put(20,10){\circle*{1.5}}
\put(12,3){{\large $N_2 = 0$}}
\put(40,10){\circle*{1.5}}
\put(35.5,3){{\large $S_2 V$}}
\put(60,10){\circle*{1.5}}
\put(48,3){{\large $N_2 = N = C V$}}
\put(20,10){\line(1,0){40}}
\end{picture}
\caption[]{Phase space in the stochastic case}
\label{fig13}
\end{figure}

\noindent
The two absorber states for the stochastic case are again:
\begin{eqnarray}
N_2&=&0\ ,\ N_1\ =\ N\ ,\\[0.3cm]
N_2&=&N\ ,\ N_1\ =\ 0\ .
\end{eqnarray}
The initial state is $N_2(t) = N_2(t=0)$:
\begin{equation}
P(N_2;0)\ =\ \delta_{N_2 N_2(t=0)}\ .
\end{equation}
The final state is:
\begin{equation}
P(N_2;t =\infty)\ =\ \sigma_1\ \delta_{0 N_2}\ +\ 
\sigma_2\ \delta_{N N_2}\ .
\end{equation}
After some calculations we get for the case $N_2(0) = 1$ (one user 
of technology $2$ occurs at $t = 0$):
\begin{equation}
\sigma_2\ =\ \frac{1}{\displaystyle \left(1 + \frac{b_1}{b_2}\right)^{N-1}}
\end{equation}
$\sigma_2$ is the probability, that the new technology wins the 
competition process.

\begin{figure}[h]
\centering
\includegraphics[width=.9\textwidth]{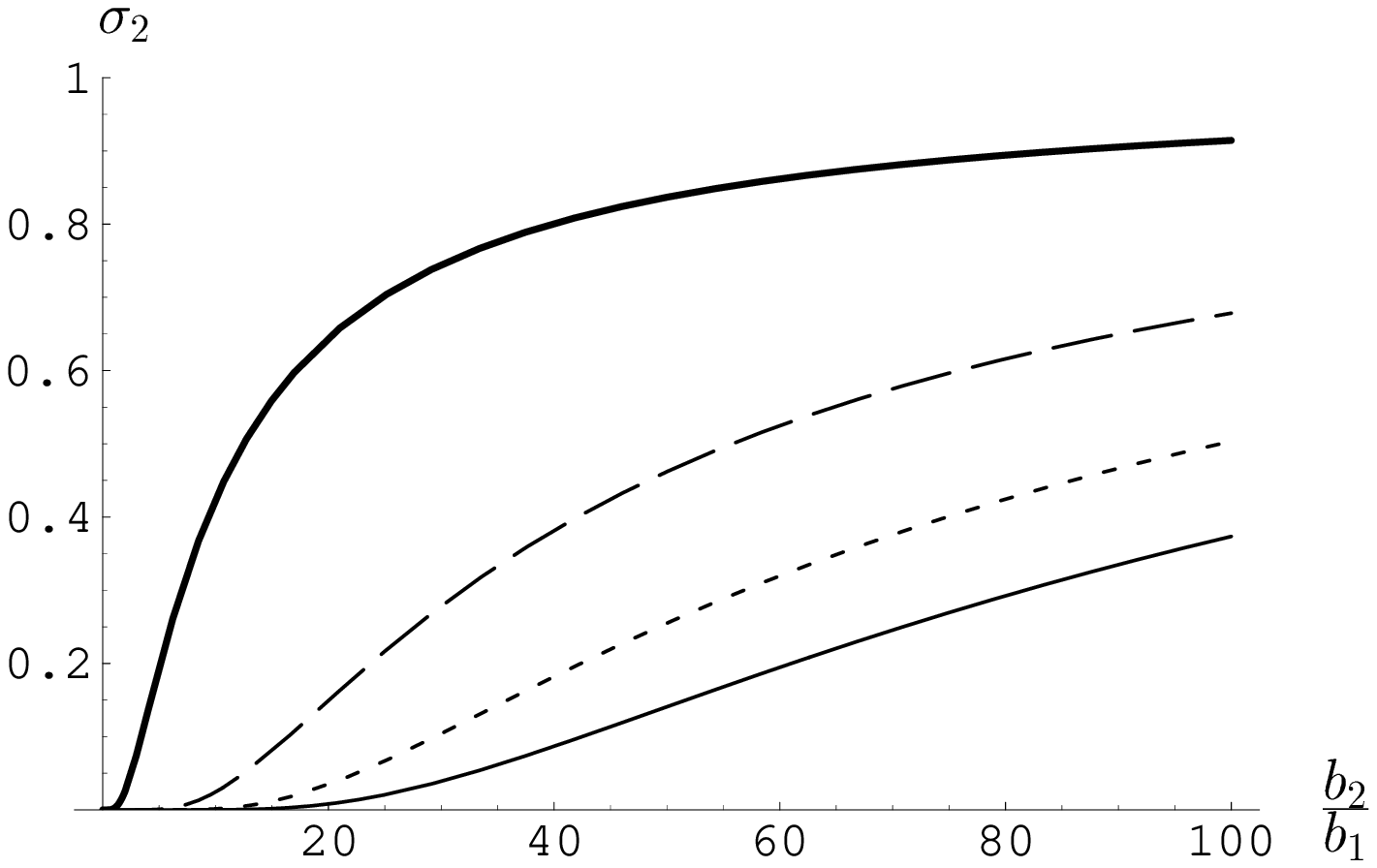}
\caption[]{Survival probability of a new technology in systems of 
different size (fat line $N=10$, dashed line $N=40$, dotted 
line $N=70$, fine line $N=100$)}
\label{fig14}
\end{figure}

\noindent
In the stochastic case, the separatrix $S$ is penetrated with a certain 
probability. The  once-forever  behaviour disappears. The  better  
technology can win the selection process with certain probability  
even if it starts only with one user at the beginning. This probability 
increases rapidly for small overall numbers of users, as follows from the 
equation and as it can be seen in Fig.\ \ref{fig14}\ \cite{ebel1986bio}.
In the stochastic description, the presence of fluctuations is responsible 
to help the technology to cross the barrier for entry in the market. 

This effect 
is size dependent. In smaller systems the probability for survival of 
new technologies even in a hyperselection situation increases. 
This can be also interpreted in economic terms. A small size of the 
system refers to the size of the market in which the competition between
the old and the new technology takes place. If a new technology is 
protected by certain mechanisms it can win the process in a
market of limited competition. This way it can obtain a significant 
number of early adopters and then go to a more open and bigger market 
with an improved chance of survival. In other words, in small niches, 
even in a lock-in situation, a new technology can replace the old one. 
By travelling from one niche to another or in growing niches 
(dynamic niche), the new technology would even be able to get control 
over the whole market. There is some empirical evidence that such 
processes take place in technological change (for a discussion see
\cite{bruc1998}).

\subsubsection{Technologies with Mixed Growth Rates}

In the following, we consider the general case where the growth rates 
of a certain technology contain both linear as well as non-linear 
terms. The transition probabilities for this case are:
\begin{eqnarray}
W^+(N_1)&=&\frac{E_1}{N}\ N_1\ N_2\ +\ \frac{b_1}{N V}\ N_2\ N_1^2\\[0.3cm]
W^-(N_1)&=& \frac{E_2}{N}\ N_1\ N_2\ +\ \frac{b_2}{N V}\ N_1\ N_2^2
\end{eqnarray}
Again here we assume that $E_{12} = E_{21} = 0$, so we
have absorber states. Further, the number of elements (plants) in 
the system is constant: $N_1 + N_2 = N = {\rm const.}$.
In analogy to the case studies above we can write the master equation:
\begin{eqnarray}
\frac{\partial}{\partial t}P(N_1;t)
&=&W^+(N_1 - 1)\ P(N_1 - 1;t)
\ +\ W^-(N_1 + 1)\ P(N_1 + 1;t)\nonumber\\[0.3cm]
&&\ -\ \left[ W^+(N_1) + W^-(N_1)\right]\ P(N_1;t)
\end{eqnarray}
In the same way as describes above we can derive the deterministic 
counterpart to the stochastic dynamics.

For the general case the survival probability for a new technology  
$\sigma_{N_2(0),N}$ can be calculated:
\begin{equation}
\sigma_{N_2(0),N}\ =\ 
\frac{\displaystyle 1 + \sum_{j=1}^{N_2(0)-1}
\prod_{i=1}^j\ 
\frac{\displaystyle E_1 + b_1 \frac{N-i}{V}}{
\displaystyle E_2 + b_2 \frac{i}{V}}
}{\displaystyle
 1 + \sum_{j=1}^{N-1}
\prod_{i=1}^j\ 
\frac{\displaystyle E_1 + b_1 \frac{N-i}{V}}{
\displaystyle E_2 + b_2 \frac{i}{V}}
}
\end{equation}
This result was first obtained in \cite{ebel1981}
and extensively applied to socio-economic problems in \cite{bruc1996b}.
As can be seen from the formula above, the survival probability for a 
new technology $\sigma$ not only depends on the selection advantage 
(the relation of the parameters $E_i$, $b_i$), but also on the size
of the system $N$ (the overall user number) and the initial number of users.

\subsubsection{Summary -- General Aspects of the Survival Probability 
of a New Technology}

The stochastic analysis of competition processes of technologies in 
a market where the number of overall users is restricted shows 
remarkable differences to the results known from a deterministic 
description. Let us note here that with the restriction to markets 
with a fixed number of economic agents, we restrict ourselves to
technological substitution processes. However, using the approach 
of under-occupied systems this restriction does not represent a limitation
for the innovation process. A variety of future possible innovative types 
is included in the system all times. The population of agents (fixed 
in size) can unlimited travel over the field of innovations. 
Already with this 
restriction the stochastic analysis compared with the deterministic one 
represents an increase in 
mathematical complexity at the descriptive level. In the multidimensional general 
case an 
analytical solution of the master equation is not available. 
Some results can be obtained using computer simulations 
\cite{schi1981,hein1981,bruc1990}.
However, what we showed so far is how statements about the 
survival probability in the long run can be derived.
Results are available for certain special cases as we presented 
above. Let us summarize the results for the special case of 
two-dimensional systems with $N_1 + N_2 = N = {\rm const.}$ and 
with the two absorber states $N_2 = 0$, $N_2 = N$. 
Absorber states means here that once the states $N_2 = 0$ or $N_2 = N$
are reached, they cannot be left anymore.

As mentioned above, this case can be treated exactly 
\cite{schi1981,ebel1981,ebel1982}.
According to the absorber character we may assume that the 
stationary probability which is  the target of evolution has a 
delta-character and can be written as:
\begin{equation}
P(N_2;t =\infty)\ =\ \sigma\ \delta_{N N_2}\ +\ 
(1-\sigma)\ \delta_{0 N_2}\ .
\end{equation}
Here, $\sigma$ is the survival probability of the second technology 
which is supposed to be the new one entering the market. An expression 
for the survival probability $\sigma$ can be calculated with the 
help of one constant of motion for the general case 
\cite{ebel1981}:
\begin{equation}
\sigma_{N_2(0),N}\ =\ 
\frac{\displaystyle 1 + \sum_{j=1}^{N_2(0)-1}
\prod_{i=1}^j\ 
\frac{W_i^-}{W_i^+}
}{\displaystyle
 1 + \sum_{j=1}^{N-1}
\prod_{i=1}^j\ 
\frac{W_i^-}{W_i^+}
}\ \ \ {\rm for}\ \ 0\ <\ N_2(0)\ <\ N
\end{equation}
and
\begin{equation}
\sigma_{N_2(0),N}\ =\ 1\ \ \ {\rm for}\ \ N_2(0)\ =\ N\ .
\end{equation}
As we introduced in subsection \ref{sec3.2.1}\ 
the transition probabilities $W^+$ and $W^-$
represent quite a variety of different processes which all are 
linked to certain  parameters. Further, the initial conditions and 
the system size N  are important for the survival of a new technology 
(see for details \cite{bruc1996b}).

In particular we considered the following representations of transition 
probabilities:

\begin{enumerate}
\item
For technologies with linear growth rates (linear case)
\begin{equation}
W_{N_2}^+\ =\ E_2\ \frac{N-N_2}{N}\ N_2\ ;
W_{N_2}^-\ =\ E_1\ \frac{N-N_2}{N}\ N_2
\end{equation}

\item
For technologies with quadratic growth rates (quadratic case)
\begin{equation}
W_{N_2}^+\ =\ b_2\ \frac{N-N_2}{N V}\ N_2^2\ ;
W_{N_2}^-\ =\ b_1\ \frac{(N-N_2)^2}{N V}\ N_2
\end{equation}

\item
For technologies with certain mixed growth rates (general case)
\begin{eqnarray}
W_{N_2}^+&=&E_2\ \frac{N-N_2}{N}\ N_2\ +\ b_2\ \frac{N-N_2}{N V}\ N_2^2
\\[0.3cm] 
W_{N_2}^-&=&E_1\ \frac{N-N_2}{N}\ N_2\ +\ b_1\ \frac{(N-N_2)^2}{N V}\ N_2
\end{eqnarray}
\end{enumerate}

We derived certain expressions for the survival probabilities by 
solving the absorber problem. These expressions can be treated 
further and certain special cases can be discusses.

\begin{enumerate}
\item
For the linear case the survival probability has been defined as
\begin{equation}
\sigma_{N_2(0),N}\ =\ 
\frac{\displaystyle 1 - \left(\frac{E_1}{E_2}\right)^{N_2(0)}
}{\displaystyle
 1 - \left(\frac{E_1}{E_2}\right)^{N_1)}}
\end{equation}
We now consider the special case of very large systems (many possible users). 
In this case $\sigma$ can be simplified to:
\begin{equation}
\sigma_{N_2(0),N\to\infty}\ =\ 0\ \ \ \ {\rm for}\ \ E_2\ <\ E_1
\end{equation}
and
\begin{equation}
\sigma_{N_2(0),N\to\infty}\ =\ 1\ -\ \left(\frac{E_1}{E_2}\right)^{N_2(0)}
\ \ \ \ {\rm for}\ \ E_2\ >\ E_1\ .
\end{equation}
If we then consider a certain initial condition, the case that a 
technology starts with just one user (plant) $N_2(0) = 1$, we 
can reduce the expression for the survival probability one step further
\begin{eqnarray}
\sigma_{N_2(0)=1,N\to\infty}&=&0\hspace{1.5cm} {\rm for}\ \ E_2\ <\ E_1\\[0.3cm]
\sigma_{N_2(0)=1,N\to\infty}&=&1\ -\ \frac{E_1}{E_2}
\ \ \ \ {\rm for}\ \ E_2\ >\ E_1\ .
\end{eqnarray}

\item
For the quadratic case the survival probability has been defined as
\begin{equation}
\sigma_{N_2(0),N}\ =\ 
\frac{\displaystyle 1 + \sum_{j=1}^{N_2(0)-1}\
\left(\frac{b_1}{b_2}\right)^j\ {N-1\choose j}
}{\displaystyle
 \left(1 + \frac{b_1}{b_2}\right)^{N-1}}
\end{equation}
This equation can be written down more specified if we consider 
a new technology starting with just one user $N_2(0)$:
\begin{equation}
\sigma_{N_2(0),N}\ =\ 
\frac{\displaystyle 1}{\displaystyle
 \left(1 + \frac{b_1}{b_2}\right)^{N-1}}
\end{equation}

\item
In the general case the survival probability has the following form.
\begin{equation}
\sigma_{N_2(0),N}\ =\ 
\frac{\displaystyle 1 + \sum_{j=1}^{N_2(0)-1}
\prod_{i=1}^j\ 
\frac{\displaystyle E_1 + b_1 \frac{N-i}{V}}{
\displaystyle E_2 + b_2 \frac{i}{V}}
}{\displaystyle
 1 + \sum_{j=1}^{N-1}
\prod_{i=1}^j\ 
\frac{\displaystyle E_1 + b_1 \frac{N-i}{V}}{
\displaystyle E_2 + b_2 \frac{i}{V}}
}
\end{equation}
In the special case of a technology starting 
with just one user $N_2(0) = 1$ we obtain:
\begin{equation}
\sigma_{N_2(0),N}\ =\ 
\frac{\displaystyle 1}{\displaystyle
 1 + \sum_{j=1}^{N-1}
\prod_{i=1}^j\ 
\frac{\displaystyle E_1 + b_1 \frac{N-i}{V}}{
\displaystyle E_2 + b_2 \frac{i}{V}}
}
\end{equation}
\end{enumerate}

In general, the transition probabilities depend on the system size, 
the system parameters and the initial conditions. In the stochastic 
case we obviously find a niche effect. In the niche, the sharpness of 
selection is diminished. In the linear case,  good  and  bad  
technologies can temporally coexist. In the quadratic case,
the  once-forever  effect of a competition and selection process 
is countered by the niche effect. The niche represents a possibility 
for a  better  technology to win the competition 
even in the situation that a lock-in effect 
can be observed. Locally developed niches may play a constructive 
role in the technological evolution and can be observed in
large complex systems, such as economic systems. 
The description, with the help of 
the master equation, includes processes, which play a role in 
these small domains (niches). In a niche, the new technology is protected 
against extinction for a limited time. After winning the 
competition in this small area, the new technology can infect the 
whole system, and may be established at the end.

\subsection{Stochastic Analysis of Multiple Decision Processes -- 
the Modelling of Technological Networks}

\subsubsection{Ehrenfest's Urn-Models with Higher Correlations}

In subsection \ref{sec3.1.1}\ 
we introduced the urn-model of Ehrenfest.The idea of
the Ehrenfest urn model was extended towards processes which
also change the total number of elements in the system. In the last 
subsections we restricted ourselves to the Ehrenfest approach by considering
transition probabilities only, and derived analytical results for the
survival probability for the special case of two competing technologies. In
this section we use again the idea of a system with constant size, but extend
the dimensionality of the system. Here, we have in mind under-occupied systems
where the number of possible different technologies is large. Only the 
number of searching plants is restricted. Now we will consider the competition
between s different technologies. This corresponds to the existence of $s$
urns, which are filled with $N_1,N_2,\ldots,N_s$ spheres. We start a
stochastic game where, at random times, spheres are taken out of an urn and put
to another urn. Here we consider a game with binary decisions, where the
transition from one urn $j$ to another urn $i$ is given by the following
transition probabilities.
\begin{eqnarray}
&&\hspace{-1cm}W(N_1\ldots,N_i + 1,\ldots,N_j-1,\ldots,N_s\vert 
N_1,\ldots,N_i,\ldots,N_j,\ldots,N_s)\nonumber\\[0.3cm]
&=&A_{ij}\ N_j\ +\ B_{ij}\ N_i N_j\ +\  
\sum_k C_{ijk}\ N_i N_j N_k
\end{eqnarray}
This is a generalization of the transition rates introduced so far in the
case of two technologies. In particular, we have in mind the mutual support
or hindering of technologies. As discussed in the economic
literature, innovations are often the outcome of an innovation network with
different actors involved. We can assume that the decision of a certain
plant or firm to implement a certain technology depends both on other firms
using the same technologies and also on other firms using different,
but some how related, technologies. The stochastic game we propose here is a
hopping process of plants between technologies.

The mean values of the occupation numbers in the thermodynamic limit
approximately follows a differential equation. In the sections above we
showed such approximation for the case of two technologies with linear,
quadratic and specific non-linear growth rates. The deterministic equations
we obtained correspond to Fisher-Eigen-Schuster equations. For the
generalization proposed above a set of Lotka-Volterra equation follows in
the deterministic limit:
\begin{eqnarray}
\frac{\rm d}{{\rm d} t} x_i&=&\sum_j\left(
A_{ij}\ x_j\ +\ B_{ij}\ N x_i x_j\ +\ \sum_k\ C_{ijk}\ N^2 x_i x_j x_k\right)
\\[0.3cm]
&&\hspace{2cm} {\rm with}\ \ \ \frac{\langle N_i\rangle}{N}\ =\ x_i\ .
\nonumber
\end{eqnarray}
The general case can only be handled by computer simulations. The equation
above comprises the case of two technologies with non-linear growth rates if
we introduce the following correspondence between parameters 
\cite{ebel2000physica}:
\begin{eqnarray}
A_{21}&=&0\ ;\ \ B_{12}\ =\ \frac{E_1}{N}\ ;\ \
C_{12k}\ =\ \frac{b_1}{N}\ \delta_{1k}\ ;\ \ k\ =\ \{1;2\}\\[0.3cm]
A_{12}&=&0\ ;\ \ B_{21}\ =\ \frac{E_2}{N}\ ;\ \
C_{21k}\ =\ \frac{b_2}{N}\ \delta_{2k}\\[0.3cm]
&&\hspace{-0.3cm}W(N_2 + 1,N_1-1\vert N_1,N_2)\ =\ E_2\ \frac{N_1 N_2}{N}\ +\
b_2\ \frac{N_2^2 N_1}{N}
\end{eqnarray}

\subsubsection{Decision Processes and the Dynamics of a Network of
Technologies}

In this section we return to the general dynamics of interacting
technologies. In subsection \ref{sec3.2.1} we gave a short economic
interpretation of processes like spontaneous generation, self-reproduction,
decay and conversion or transition. All these processes can be interpreted
in terms of decision processes by firms or plants related to expansion or
shrink and related to the choice of different technologies from a set of
technologies available. The case of innovation interpreted as first
occupation of a so far unoccupied technological possibility is just one
process in a whole set of decision processes made by firms. The substitution
case between old and new technologies that we discussed earlier in detail
represents a very specific decision. In the following, we will give a detailed
interpretation of different possible decision processes (for further
economic interpretation please consult also \cite{bruc1996b}). 

We start again with the set of occupation numbers. $N_i$ is
the number of plants using a certain technology $i$. We consider a network of
$s$ different technologies competing with each other  $i = 1,\ldots,s$. We
will no longer stick to the assumption that the total number of plants in
the system will remain constant. This way, we also consider growth and
decline processes not only of certain used technologies in the system but
also of the system, the market as a whole.

We can differentiate between changes which take place inside one type of
users (where the type or the group that the users belong to is characterized by
the technology they use). For instance, the number of users of the
technology $i$ may increase or decrease. This is a stepwise process which only
changes the occupation number by one:
\begin{equation}
\begin{array}{c@{\ \ \ \longrightarrow\ \ \ }c}
(N_i)&(N_i + 1)\\[0.3cm]
(N_i)&(N_i - 1)
\end{array}
\end{equation}

Furthermore, changes can take place between different types of users.
Mathematically this is expressed by the simultaneous change of two
occupation numbers:
\begin{equation}
{N_i\choose N_j}\ \ \longrightarrow\ \ {N_i-1\choose N_j+1}
\end{equation}
Plants can take decisions to leave a certain technology they have been using
before, they can develop a new technology, or use a technology already
established in the market. They can also develop further a technology 
they have already used and in this way create a new type of technology. 
Not only technologies
can enter the market; firms also can enter the market. The creation of a new
firm might be connected to a established technology, related to the
introduction of a new technology. The problem technologies face by their
introduction into a market are comparable to the problems firms have by
entering a certain market. Here barriers for entry might also occur which
hinder a certain technology in entering a market. On the other side, a network
of firms might create support for a new firm, or also for a new technology, to enter
the market (the system). Coalitions and cooperation are examples of
synergetic effects in the introduction of new technologies.

We will now formulate some possible changes in the language of transition
probabilities:

\noindent
\begin{enumerate}
\item
The number of plants using a certain technology $i$ increases:
\begin{equation}
N_i\ \longrightarrow\ N_i +1
\end{equation}
This change can be the result of different processes which are further
differentiable.

\begin{itemize}
\item
$A_i^0 N_i$ -- linear self-reproduction: the number of plants will
grow according to the existing number of plants. If these plants
belong to one firm, firm growth is modelled. If they belong to different
firms, the growth of an industrial sector is considered. The growth rate of
this process is linear. Let us remember here that if only this process 
takes place in the system the technology would increase exponentially.

\item
$A_i^1 N_i^2$ -- self-amplification (second order
self-reproduction): here the number of plants already using the technology
creates a network effect of higher order which speeds up the growth momentum
for this specific technology.

\item
$B_{ij}\ N_i N_j$ -- sponsoring or supporting from other plants: in this
case plants using a different technology $j$ are relevant for the growth of
the technology $i$. One can think of systems of coupled technologies, one
supporting the other, or of production chains where one technology relies
on others.
\end{itemize}
For this case the transition probabilities can be written:
\begin{equation}
W(N_i + 1,N_j\vert N_i,N_j)\ =\ A_i^0\ N_i\ +\ A_i^1\ N_i^2\ +\ 
B_{ij}\ N_i N_j
\end{equation}
The different coefficients or parameters describe the strength of a
certain effect which is acting in the system.

\item
Spontaneous formation of a new plant with a certain technology
\begin{equation}
N_i\ \longrightarrow\ N_i +1
\end{equation}

\begin{itemize}
\item
$\Phi_0$ -- spontaneous formation: here the entry of plants is
not connected to the number of those already existing. One can think of a
startup. The startup can either start with an already existing technology or
be linked to the development of a new technology. In the latter case, $N_i$
would be zero at the beginning. One would usually assume that the number of
new plants created spontaneously is relatively low.
\end{itemize}

\begin{equation}
W(N_i + 1,N_j\vert N_i,N_j)\ =\ \Phi_0
\end{equation}

\item 
Decrease of the number of the plants using technology $i$
\begin{equation}
N_i\ \longrightarrow\ N_i -1
\end{equation}
\begin{itemize}
\item
$D_i^0 N_i$ -- linear decrease: Given that each technology in the
market occupies a certain niche in the market, one can assume that the number
of plants using a technology which can not survive will depend on the size
of the population of all plants using that technology.

\item
$D_i^1 N_i^2$ -- non-linear decrease or restricted capacity: this
process remains for a network effect of the number of plants using a certain
technology which drives plants out of the market. This process is
responsible for the existence of a restricted capacity in markets with
linear growing technologies. Without the existence of decrease terms of
higher order, one would be confronted with infinite, exploding markets which
stay in contradiction to empirical observations.
\end{itemize}

\begin{equation}
W(N_i - 1,N_j\vert N_i,N_j)\ =\ D_i^0\ N_i\ +\ D_i^1\ N_i^2
\end{equation}

\item
Origin of a new technology connected to the formation of a new plant 
(induced innovation)
\begin{equation}
{0\choose N_j}\ \ \longrightarrow\ \ {N_i = 1\choose N_j}
\end{equation}

\begin{itemize}
\item
$M_{ij} N_j$-- Here we assume that the creation of a startup with
a new technology is not a pure spontaneous process but is related to the number
of plants using another (relevant for the new technology) technology.
\end{itemize}

\begin{equation}
W(N_i = 1,N_j\vert N_i = 0,N_j)\ =\ M_{ij}\ N_j
\end{equation}

\item
Change in the use of a technology (conversion, transition)
\begin{equation}
{N_i\choose N_j}\ \ \longrightarrow\ \ {N_i + 1\choose N_j -1}
\end{equation}

\begin{itemize}
\item
$A_{ij} N_j$ -- simple transition from $j$ to $i$: in this case the
decision to take over a new technology is only influenced by the number of
plants using a certain technology. One can interpret this process in the
following way. If the number of plants using the same technology increases, then
the competition between these plants also increases and plants might be
motivated to look for another technology to increase their chances on the
market. In the case where the technology is not yet occupied (not yet
invented), the transition will also create an innovation for the system. Let
us note here that beside spontaneous generation and the induced innovation,
this process is important for the exploration of new areas in the
technological space.

\item
$B_{ij} N_i N_j$ -- transition from $j$ to $i$, in addition 
is promoted by $i$: 
in general one can assume that plants do not act in isolation. Conversely, 
the information flows between firms and plants about market
conditions and technological change, are an important part of economic
processes. Important in the process we discuss here is the decision to use a certain
new technology $j$ (here we use {\it new} in the sense that the technology is new
for the plant) which is related to the number of firms already using this technology.
We can further assume that the number of firms using a technology can be
interpreted as a measure of attractiveness of this particular technology.
In this case, the process represents one form in which imitation can be
modelled.

\item
$C_{ijk} N_i N_j N_k$ --  
transition from $j$ to $i$, that in addition is promoted
(sponsored) by $j$ and $k$: 
this process represents one possibility to introduce a
network effect in the decision process of one firm to use a certain
technology. One can imagine that the technologies $j$ and $k$ are related in
sense of a production chain or that they complement each other.
\end{itemize}

\begin{eqnarray}
&&\hspace{-2.5cm}W(N_i + 1,N_j - 1\vert N_i,N_j)\nonumber\\[0.3cm]
&=& A_{ij}\ N_j\ + \ B_{ij}\ N_i N_j\ +\ \sum_k C_{ijk}\ N_i N_j N_k
\end{eqnarray}
\end{enumerate}

The introduced model is composed in a modular way. Different processes 
related to the change in the occupation number space have been introduced.
We also called these processes elsewhere elementary processes. In the
presentation above we tried to give examples of processes relevant in the
decision behaviour of firms using different technologies. However,
alternative definition and the introduction of further processes are
possible within the model framework. The task consists of the definition of
processes which can be observed empirically in the economy. The model
represents a specific way to operationalize processes of decision making
inside firms, the information flows between firms and the interactive
pattern between technologies. By relating all these processes to the
occupation number space a certain reduction of information takes place. On
the other hand, the different parameters allows for the possibility of including
further economic information. The advantage of the modular structure of the
model is that it puts different processes together and places them in an
evolutionary framework. Growth of firms, substitution processes, invention
and imitation, startups and firm closing are all part of one model. If we
link the model to its deterministic counterpart, the instrumentarium of
dynamic systems becomes available. This way, at least, we can hope to get some
insights in the analytic structure of the model and possible stationary
states as well as their stability behaviour. Using the stochastic model for
simulations we can obtain a lot of statements about the systems behaviour.
Interesting investigations can be made by variation of the parameters.
Different kinds of connections can be analyzed.

Beyond the economic interpretation chosen for the model, the framework can be
also applied in quite different contexts. In any of these new application areas
types, elements and the 
network of connections have to be interpreted at new again. Some of the
authors applied the model for biochemical processes \cite{ebel1986bio}, 
for growth, competition and evolution of scientific specialties 
\cite{bruc1990}, and for the dynamics of values and competences 
\cite{scha1999}. 

\subsubsection{Further Analysis of the Probability Distribution }

In the case that we obtain the probability distribution 
$P(N_1,\ldots,N_s;t)$ 
analytically, or by computer simulations, it is possible to get
the time-behaviour of the moments (mean values, correlation functions) in
dependence on the parameters of the system using the generation function
\cite{hein1981,ebel1986bio}. For the partial
differential equation given in section \ref{sec3.2.1}\ 
we can get an approximate
solution by introducing the transformation \cite{nico1977} 
\begin{equation}
F\ =\ \exp\left[\Psi(\eta_i) N\right]\ ;\ \ \ \eta_i\ =\ s_i\ -\ 1
\end{equation}
at which the function $\psi(\eta_i)$ is expanded in a Taylor series
\begin{equation}
\Psi\ =\ \sum_i\ a_i\ \eta_i\ +\ 
\frac{1}{2!}\sum_{i,j}\ b_{ij}\ \eta_i \eta_j
\ +\ 
\frac{1}{3!}\sum_{i,j,k}\ c_{ijk}\ \eta_i \eta_j \eta_k\ +\ \ldots
\end{equation}
The coefficients $a_i$ and $b_{ij}$ are in terms of
the moments of the probability distribution as follows:
\begin{eqnarray}
a_i&=&\frac{\langle N_i\rangle}{N}\\[0.3cm]
b_{ii}&=&\frac{1}{N}\ \left[ \left\langle N_i^2\right\rangle
- \langle N_i\rangle\right]\\[0.3cm]
b_{ij}&=&\left[ \langle N_i N_j\rangle\right]
\end{eqnarray}
These equations give us information about the deviation from the
Poisson-distribution (Poisson-distribution means for the coefficients
$b_{ij} = 0,\ \forall\ i,j$ 
and all exponential coefficients of higher order are also
zero). From the equations we can get a closed set of differential equations
for the mean values $a_i$ and the variances $b_{ij}$. 
Provided that we can neglect the coefficients higher than
second order, we can achieve an approximative solution of the set of
differential equations. Especially for processes which satisfy a
multi-Poisson distribution, the coefficients higher than first order are
zero in the Taylor expansion. This is the case if we get for $F$ or $\Psi$, 
respectively, a partial differential equation linear in
$s_i$ and at most of first order. If we know in this case the
mean values $a_i = \langle N_i\rangle /N$ 
and start at $t = 0$ with a
multi-Poisson distribution, we can get an exact solution for the probability
distribution. The solution is a multi-Poisson distribution for all times
\begin{equation}
P(N;t)\ =\ \prod_i\ 
\frac{\displaystyle \langle N_i(t)\rangle^{N_i}}{\displaystyle N_i !}\
\exp\left[- \langle N_i\rangle\right]
\end{equation}
Now we answer the question, which processes (transition probabilities from
subsection \ref{sec3.2.1}) 
lead to a multi-Poisson distribution, i.e., which processes have in
the generating function at most terms with a first derivative with respect
to $s_i$ and are linear in $s_i$. Then, the
generating function looks as follows:
\begin{equation}
\dot{F}(s;t)\ =\ \sum_j\ A_i^{(2)}\ (1 - s_i)\ 
\frac{\partial F}{\partial s_i}\ +\ A_{ij}^{(3)}\ (s_i - s_j)\ 
\frac{\partial F}{\partial s_j}
\end{equation}
So we can get
\begin{eqnarray}
\dot{\Psi}(\eta;t)&=&\sum_i \sum_j\ \left\{ - A_i^{(2)}\ \eta_i a_i\ +\
A_{ij}^{(3)}\ (\eta_i - \eta_j)\right\}\\[0.3cm]
\dot{a}_i&=&- A_i^{(2)}\ a_i\ +\ \sum_{i\ne j}\ A_{ij}^{(3)}\ a_j
\ -\ a_i\ \sum_{i\ne j}\ A_{ij}^{(3)}
\end{eqnarray}
If we start at $t = 0$ with a multi-Poisson distribution, the
probability distribution remains a multi-Poisson distribution for 
all times
\begin{equation}
P(N;t)\ =\ 
\exp\left[- \langle N(t)\rangle_C\right]\ 
\prod_i\ 
\frac{\displaystyle\langle N_i(t)\rangle^{N_i}}{\displaystyle N_i !}
\end{equation}
The extinction probability of the whole component is
\begin{equation}
P(0;t)\ =\ 
\exp\left[- \langle N_C(t)\rangle\right]
\end{equation}
This extinction probability depends on the overall number of individuals
$N_C$ of the component and decreases exponentially with 
$\langle N_C(t)\rangle$.

A detailed description for the calculation of the moments of the probability
distribution $P(N_1,\ldots,N_s;t)$ and the probability distribution was given
in the work of Heinrich and Sonntag \cite{hein1981}.
The time-behaviour of the
moments could be achieved with the help of generating function. Deviations
 from the Poisson behaviour were investigated. Fluctuations and their
influence on the system structure and behaviour could be analyzed. The
analysis of time-dependent correlation functions could also be interesting
for the description of socio-economic systems.

\section{Summary}

This paper is devoted to the study of innovation processes in socio-economic
contexts. In particular, we investigate the influence of stochastic effects
on processes of self-organization and evolution. We take a special network
perspective. Starting with the new emergent field of complex networks theory,
we develop our own approach of sensitive networks relevant to the
description of an innovation.

Innovation is first introduced on a general level as a specific process that
changes the composition and dynamic constitution of a system. We use a
discrete representation of the system in terms of a space of occupation
numbers. Then, innovation can be described as a hopping process between
positive cones. Further, we introduce the notion of an under-occupied system.
This way we implement a set of possible future paths of developments in our
modelling. We relate this abstract notion of an innovation to the discussion
of innovation processes in economics.

In the economic literature, innovation has been understood as the outcome of
processes running on networks of different actors. 
We concentrate on firms and technologies in this paper. We present
a network theory of innovation by mapping the dynamic interactions related
to the emergence of an innovation in form of a graph.

\begin{figure}[h]
\vspace{0cm}
\centering
\includegraphics[width=.85\textwidth]{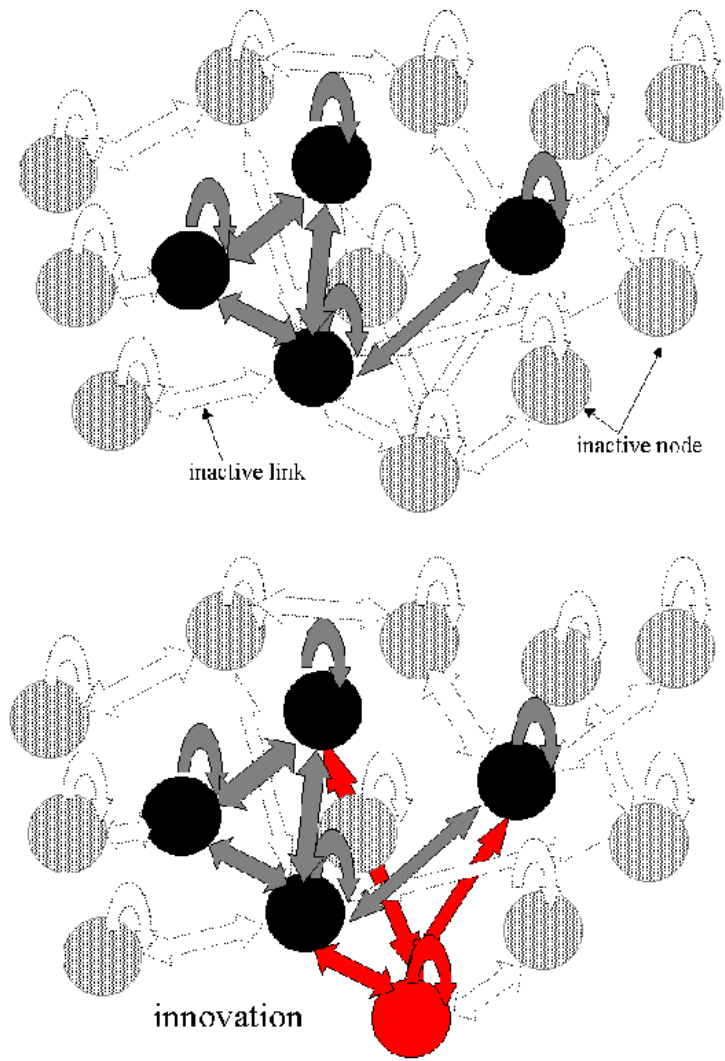}
\caption[]{Illustration of a sensitive network. Embedded 
in a large network of inactive nodes and links the activated
part of the network (upper part of the figure) 
changes its composition when an innovation is emerging (lower part of the 
figure).}
\label{fig15}
\end{figure}

Figure \ref{fig15} illustrates our approach. The system is composed of a
large set of enumerable types. Each of these types is represented by
a node. At a certain point in time, only a small part of these nodes are active.
The pattern of interaction between them (including processes of self-influence)
determines the dynamic composition of the system. It is visualized 
in terms of (active) links between the nodes. This active network will 
produce a dynamics which has a certain set of stable states. 
We assume that the activated
part of the network is embedded in a much larger network of inactive nodes 
and links. The inactive nodes represent future possibilities in the evolution
of the system. An innovation appears when a unoccupied node becomes
occupied for the first time. With this first occupation, the 
set of links connecting the "new" node with already occupied nodes also
becomes activated. It is readily apparent that such an event 
changes the whole composition of the 
system. Accordingly, the stable state that the system might have reached already becomes
instable and the system searches for a new stable state. If we assume that
the interaction between the nodes (types) is a competitive one, the stable state
of a certain activated network can also include the deactivation of certain nodes.
Types (nodes) which are selected out will transit to a non-occupied inactive status.

We first discuss structural (static) properties of such a relational
network. In this way, instruments from random graph theory and percolation theory
become relevant. In particular, formulae for the probabilities of the
occurrence and the distribution of components and cycles in large networks
have been obtained by combinatorial considerations. We show that
connectivity is a central measure in the structure of these kinds of
networks. Evolving, dynamic networks show different phenomena
compared to random graphs. The results better correspond to findings in
real data of empirical networks such as the appearance of power law
distributions of the degree function.

In the centre of the paper we turn to the discussion towards the dynamic
properties of networks. In particular, we derive descriptions for a network
of interacting technologies and interacting firms. We call these networks
sensitive because innovation processes described in terms of the removal or
appearance of a node might change  the dynamic behaviour of the
system dramatically. We use a stochastic description of such an evolving network and base
this on the theory of birth and death processes. We introduce different
forms of transition probabilities for closed systems as well as for open,
growing and declining systems. We define master equations and non-linear
differential equations as their deterministic equivalents. Contrary 
to deterministic models, the stochastic description offers the advantage that,
at finite times, new technologies (types) can arise or ``die out''. Further,
the emergence of an innovation can be treated as a singular stepwise event
implying the transition from an under-occupied to an occupied state.

In some special cases the master equation can be solved analytically and a
stationary survival probability for an innovation can be derived. We show
that the stochastic dynamics differ essentially from the deterministic.
Separatrices, which decompose the phase space, cannot be intersected in the
deterministic case. In the stochastic case they can be crossed. This way, a
``once-for-ever'' or hyperselection that is known in economics as lock-in
phenomena for technologies can be avoided.

The stochastic model of networked dynamic interactions of technologies is
further generalized in a multi-dimensional case. Processes representing
different non-linearities are discussed in the context of technological
change. Although we use technological evolution as main reference point, we
also point to the fact that the modular structure of the model also allows 
for its application in quite different fields. Applications in fields as
biology, population theory and science of science have been presented
\cite{feis1989,bruc1990}.

\clearpage
\addcontentsline{toc}{section}{Index}
\flushbottom
\printindex


\begin{thebibliography}{777}
%
\addcontentsline{toc}{section}{References}

\bibitem{albe2000prl}
Albert, R., Barab\'asi, A.-L.\ (2000) Topology of Evolving Networks: 
Local Events and Universality. Physical Review Letters 85:5234-5237
[arXiv:cond-mat/0005085]

\bibitem{albe2002}
Albert, R., Barab\'asi, A.-L.\ (2002) Statistical Mechanics of
Complex Networks. Reviews of Modern Physics 74:47-97
[arXiv:cond-mat/0106096]

\bibitem{albe1999}
Albert, R., Jeong, H., Barab\'asi, A.-L.\ (1999) Diameter of 
the World-Wide Web. Nature (London) 401:130-131
[arXiv:cond-mat/9907038]

\bibitem{albe2000nat}
Albert, R., Jeong, H., Barab\'asi, A.-L.\ (2000) Error and Attack 
Tolerance of Complex Networks. Nature (London) 406:378-382, 
409:542 (correction) [arXiv:cond-mat/0008064]

\bibitem{alle1994}
Allen, P.~M.\ (1994) Coherence, Chaos and Evolution in the Social Context. 
Futures 26:583-597 

\bibitem{bara2002new}
Barab\'asi, A.-L.\ (2002) Linked: The New Science of Networks. 
Perseus Publishing, Cambridge, MA

\bibitem{bara2002phys}
Barab\'asi, A.-L., Jeong, H., N\'eda, Z., Ravasz, E., Schubert, A.,
Vicsek, T.\ (2002) Evolution of the Social 
Network of Scientific Collaborations. Physica A 311:590-614
[arXiv:cond-mat/0104162]

\bibitem{bart1958a}
Bartholomay, A.~F.\ (1958) On the Linear Birth and Death Processes of 
Biology as Markoff Chains. The Bulletin of Mathematical Biophysics 20:97-118

\bibitem{bart1958b}
Bartholomay, A.~F.\ (1958) Stochastic Models for Chemical Reactions: 
I. Theory of the Unimolecular Reaction Process.
The Bulletin of Mathematical Biophysics 20:175-190

\bibitem{bart1959}
Bartholomay, A.~F.\ (1959)  Stochastic Models for Chemical Reactions: 
II. The Unimolecular Rate Constant. 
The Bulletin of Mathematical Biophysics 21:363-373

\bibitem{bian2001euro}
Bianconi, G., Barab\'asi, A.-L.\ (2001) Competition and Multiscaling 
in Evolving Networks. Europhysics Letters 54:436-442
[arXiv:cond-mat/0011029]

\bibitem{bian2001prl}
Bianconi, G., Barab\'asi, A.-L.\ (2001) Bose-Einstein Condensation 
in Complex Networks. Physical Review Letters 86:5632-5635 
[arXiv:cond-mat/0011224]

\bibitem{born2003}
Bornholdt, S., Schuster, H.~G.\ (Eds.) (2003) Handbook of Graphs and
Networks: From the Genome to the Internet. Wiley-VCH Verlag, Weinheim  

\bibitem{brod2000}
Broder, A., Kumar, R., Maghoul, F., Raghavan, P., Rajagopalan, S., 
Stata, R., Tomkins, A., Wiener, J.\ (2001) Graph Structure in the Web.
Computer Networks -- The International Journal of Computer and 
Telecommunications Networking 33:309-320. This article is part
of the special issue: Proceedings of the 9.\ International 
World Wide Web Conference.
The Web: the Next Generation. Amsterdam, May 15- 19, 2000.
Computer Networks -- The International Journal of Computer and 
Telecommunications Networking 33(1-6):v-viii, 1-843

\bibitem{bruc1994}
Bruckner, E., Ebeling, W., Jim\'enez-Monta\~no, M.~A., Scharnhorst, 
A.\ (1994) Hyperselection and Innovation Described by a Stochastic
Model of Technological Evolution. In: Leydesdorff, L., van den
Besselar, P.\ (Eds.) Evolutionary Economics and Chaos Theory:
New Directions in Technology Studies. Pinter Publishers, London, 
pp.\ 70-90

\bibitem{bruc1996b}
Bruckner, E., Ebeling, W., Jim\'enez-Monta\~no, M.~A., Scharnhorst, 
A.\ (1996) Nonlinear Stochastic Effects of Substitution: An 
Evolutionary Approach. Journal of Evolutionary Economics 6:1-30

\bibitem{bruc1989}
Bruckner, E., Ebeling, W., Scharnhorst, A.\ (1989) Stochastic Dynamics of 
Instabilities in Evolutionary Systems. System Dynamics Review 5:176-191

\bibitem{bruc1990}
Bruckner, E., Ebeling, W., Scharnhorst, A.\ (1990) The Application of 
Evolution Models in Scientometrics. Scientometrics 18:21-41

\bibitem{bruc1998}
Bruckner, E., Ebeling, W., Scharnhorst, A.\ (1998) Technologischer
Wandel and Innovation -- Stochastische Modelle f\"ur innovative
Ver\"anderungen in der \"Okonomie [Technological Change and
Innovation -- Stochastic Models for Innovative Changes in the 
Economy]. In: Schweitzer, F., Silverberg, G.\ (Eds.) Evolution
und Selbstorganisation in der \"Okonomie. Jahrbuch der 
Selbstorganisation, Vol.\ 9. Duncker \& Humblot, Berlin,
pp.\ 361-382 [in German]

\bibitem{buch2002}
Buchanan, M.\ (2002) Nexus: Small Worlds and the Groundbreaking Science 
of Networks. W.~W.\ Norton \& Company, New York, London

\bibitem{cast1979}
Casti, J.~L.\ (1979) Connectivity, Complexity and Catastrophe in 
Large-Scale Systems.
International Series on Applied Systems Analysis, Vol.\ 7.
John Wiley \& Sons, Chichester, New York

\bibitem{doro2002}
Dorogovtsev, S.~N., Mendes, J.~F.~F., Samukhin A.~N.\ (2003) 
Principles of Statistical Mechanics of Uncorrelated Random Networks. 
Nuclear Physics B 666:396-416 [arXiv:cond-mat/0204111]

\bibitem{dosi1988}
Dosi, G., Freeman, Ch., Nelson, R., Silverberg, G., Soete, L.\ (Eds.)
(1988) Technical Change and Economic Theory. IFIAS Research Series,
Vol.\ 6. Pinter Publishers, London

\bibitem{dros2003}
Drossel, B., McKane, A.~J.\ (2003) Modelling Food Webs. In:
Bornholdt, S., Schuster, H.~G.\ (Eds.) Handbook of Graphs and
Networks: From the Genome to the Internet. Wiley-VCH Verlag, Weinheim,
pp.\ 218-247 [arXiv:nlin.AO/0202034]

\bibitem{ebel1990phys}
Ebeling, W., Engel, A., Feistel, R.\ (1990) Physik der Evolutionsprozesse.
Akademie-Verlag, Berlin [in German]

\bibitem{ebel1990selb}
Ebeling, W., Engel, H., Herzel, H.\ (1990) Selbstorganisation in der Zeit.
Wissenschaftliche Taschenb\"ucher, Vol.\ 309. Akademie-Verlag, Berlin
[in German]

\bibitem{ebel1982}
Ebeling, W., Feistel, R.\ (1982, 2. ed.\ 1986) Physik der 
Selbstorganisation und Evolution. Akademie-Verlag, Berlin [in German]

\bibitem{ebel1994}
Ebeling, W., Feistel, R.\ (1994) Chaos und Kosmos: Prinzipien  der  Evolution.
Spektrum-Verlag, Heidelberg, Berlin, Oxford [in German]

\bibitem{ebel1998}
Ebeling, W., Freund, J., Schweitzer, F.\ (1998) Komplexe Strukturen: 
Entropie und Information. B.\ G.\ Teubner-Verlag, Stuttgart, Leipzig 
[in German]

\bibitem{ebel2001}
Ebeling, W., Karmeshu, Scharnhorst, A.\ (2001) Dynamics of Economic 
and Technological Search Processes in Complex Adaptive Landscapes. 
Advances in Complex Systems 4(1):71-88.
This article is part of the special issue: Helbing, D.,
Schweitzer, F.\ (Guest Eds.) Complex Dynamics in Economics.
Advances in Complex Systems 4(1):1-176

\bibitem{ebel2000physica}
Ebeling, W., Molgedey, L., Reimann, A.\ (2000) Stochastic Urn Models 
of Innovation and Search Dynamics. Physica A 287:599-612

\bibitem{ebel1986czech}
Ebeling, W., Scharnhorst, A.\ (1986) Selforganization Models for Field 
Mobility of Physicists. Czechoslovak Journal of Physics B 36:43-46

\bibitem{ebel2000traffic}
Ebeling, W., Scharnhorst, A.\ (2000) Evolutionary Models of Innovation 
Dynamics. In: Helbing, D., Herrmann, H.~J., Schreckenberg, M.,
Wolf, D.~E.\ (Eds.) Traffic and Granular Flow '99: Social, Traffic, 
and Granular Dynamics. Springer, Berlin, Heidelberg, New York, pp.\ 43-56

\bibitem{ebel1999}
Ebeling, W., Scharnhorst, A., Jim\'enez-Monta\~no, M.~A., Karmeshu
(1999) Evolutions- und Innovationsdynamik als Suchproze\ss\ in 
komplexen adaptiven Landschaften [Evolution and Innovation Dynamics
as Search Processes in Complex Adaptive Landscapes]. In: Mainzer, K.\ (Ed.) 
Komplexe Systeme und Nichlineare Dynamik in Natur und Gesellschaft:
Komplexit\"atsforschung in Deutschland auf dem Weg ins n\"achste 
Jahrhundert. Springer-Verlag, Berlin, Heidelberg, New York,
pp.\ 446-473 [in German]

\bibitem{ebel1986bio}
Ebeling, W., Sonntag, I.\ (1986) A Stochastic Description of Evolutionary
Processes in Underoccupied Systems. BioSystems 19:91-100

\bibitem{ebel1981}
Ebeling, W., Sonntag, I., Schimansky-Geier, L.\ (1981) On the Evolution 
of Biological Macromolecules II: Catalytic Networks. studia 
biophysica 84:87-88 and microfiche 1/37.54

\bibitem{ehre1907}
Ehrenfest, P., Ehrenfest, T.\ (1907) \"Uber zwei bekannte Einw\"ande 
gegen das Boltzmannsche H-Theorem [On two known objections against
the H-theorem of Boltzmann]. Physikalische Zeitschrift 8:311-314
[in German].
Reprinted in: Klein, M.~J.\ (Ed.) with an introduction by Casimir,
H.~B.~G.\ (1959) Paul Ehrenfest: Collected Scientific Papers. 
North-Holland Publishing, Amsterdam, item 15, pp.\ 146-149

\bibitem{eige1971}
Eigen, M.\ (1971) The Selforganization of Matter and the Evolution of 
Biological Macromolecules. Die Naturwissenschaften 58:465-523

\bibitem{eige1977}
Eigen, M., Schuster, P.\ (1977) The Hypercycle: A Principle of Natural 
Self-Organization. Part A: Emergence of the Hypercycle.
Die Naturwissenschaften 64:541-565

\bibitem{eige1978c}
Eigen, M., Schuster, P.\ (1978) The Hypercycle: A Principle of Natural 
Self-Organization. Part C: The Realistic Hypercycle.
Die Naturwissenschaften 65:341-369

\bibitem{erdo1959}
Erd\"os, P., R\'enyi, A.\ (1959) On Random Graphs I. 
Publicationes Mathematicae (Debrecen) 6:290-297.
Reprinted in \cite{tura1976}, item 158, pp.\ 308-315

\bibitem{erdo1960}
Erd\"os, P., R\'enyi, A.\ (1960) On the Evolution of Random Graphs. 
A Magyar Tudom\'anyos Akad\'emia Matematikai \'es Fizikai 
Tudom\'anyok Oszt\'aly\'anak K\"ozlem\'enyei    
[Publications of the Mathematical Institute of the Hungarian Academy of
Sciences] 5:17-61.
Reprinted in \cite{tura1976}, item 172, pp.\ 482-525

\bibitem{falo1999}
Faloutsos, M., Faloutsos, P., Faloutsos, C.\ (1999) On Power-Law
Relationships of the Internet Topology.
Computer Communications Review 29:251-262

\bibitem{feis1977}
Feistel, R., Ebeling, W.\ (1978) Deterministic and Stochastic Theory 
of Sustained Oscillations in Autocatalytic Reaction Systems.
Physica A 93:114-137

\bibitem{feis1989}
Feistel, R., Ebeling, W.\ (1989) Evolution of Complex Systems.
Physikalische Monographien. VEB Deutscher Verlag der Wissenschaften,
Berlin; Evolution of Complex Systems: Selforganisation, Entropy
and Development. Mathematics and Its Applications, East European
Series, Vol.\ 30. Kluwer Academic Publishers, Dordrecht, Boston,
London

\bibitem{fell1997}
Fell, D.\ (1997) Understanding the Control of Metabolism.
Frontiers in Metabolism, Vol.\ 2. Portland Press, London

\bibitem{fell2000}
Fell, D., Wagner, A.\ (2000) The Small World of Metabolism. Nature
Biotechnology 18:1121-1122

\bibitem{fell1951}
Feller, W.\ (1951) Two Singular Diffusion Problems.
Annals of Mathematics, 2. Ser.\ 54:173-182

\bibitem{fran1994}
Frankhauser, P.\ (1994)  La Fractalit\'e des Structures [sic] Urbaines.
Collection Villes. \'Edition Anthropos, Paris [in French]

\bibitem{fren2000}
Frenken, K.\ (2000) A Complexity Approach to Innovation Networks: 
The Case of the Aircraft Industry (1909-1997). 
Research Policy  -- A Journal Devoted to Research Policy,
Research Management and Planning 29:257-272

\bibitem{gard1970}
Gardner, R.~M., Ahsby, W.~R.\ (1970) Connectance of Large Dynamic 
(Cybernetic) Systems: Critical Values for Stability. 
Nature (London) 228:784

\bibitem{hake1978}
Haken, H.\ (1978) Synergetics: An Introduction, 2.\ enl.\ ed.,
(1983) 3.\ rev.\ enl.\ ed.. Springer Series in Synergetics, Vol.\ 1.
Springer-Verlag, Berlin, Heidelberg, New York

\bibitem{hake1988inf}
Haken, H.\ (1988) Information and  Selforganization: A Macroscopic 
Approach to Complex Systems. Springer Series in Synergetics, Vol.\ 40.
Springer-Verlag, Berlin, Heidelberg, New York 

\bibitem{hake1996}
Haken, H.\ (1996) Synergetik und Sozialwissenschaften
[Synergetics and Social Sciences].
Ethik und Sozialwissenschaften: EuS -- Streitforum f\"ur 
Erw\"agungskultur 7:587-594 [in German]

\bibitem{hara1965}
Harary, F., Norman, R.~Z., Cartwright, D.\ (1965) Structural Models:
An Introduction to the Theory of Directed Graphs.
John Wiley \& Sons, New York, London, Sydney 

\bibitem{hayt1996}
Haythornthwaite, C.\ (1996) Social Network Analysis: An Approach 
and Technique for the Study of Information Exchange.
Library and Information Science Research 18:323-342

\bibitem{hein1981}
Heinrich, R., Sonntag, I.\ (1981) Analysis of the Selection 
Equations for a Multivariable Population Model: Deterministic 
and Stochastic Solutions and Discussion of the Approach for 
Populations of Self-Reproducing Biochemical Networks.
Journal of Theoretical Biology 93:325-361

\bibitem{helb1997}
Helbing, D.\ (1997) Traffic Dynamics: New Physical Modeling Concepts.
Springer-Verlag, Berlin, Heidelberg, New York

\bibitem{hube2001}
Huberman, B.~A.\ (2001) The Laws of the Web: Patterns in the 
Ecology of Information. The MIT Press, Cambridge, MA

\bibitem{hunt2000}
Hunter, L., Elias, M.~J.\ (2000) Interracial Friendships, 
Multicultural Sensitivity, and Social Competence: How Are 
They Related? Journal of Applied Developmental Psychology 20:551-573

\bibitem{jack2001}
Jackson, M.~O., van den Nouweland, A.\ (2001) Strongly Stable Networks.
Manuscript, June 2001, revised May 2003. Available at the URL's:
http://www.hss.caltech.edu/$\sim$jacksonm/coopnet.pdf,\hfill

http://www.hss.caltech.edu/$\sim$jacksonm/coopnet.ps.\hfill

Games and Economic Behavior, forthcoming

\bibitem{jain2001}
Jain, S., Krishna, S.\ (2001) A Model for the Emergence of
Cooperation, Interdependence and Structure in Evolving Networks.
Proceedings of the National Academy of Sciences of the United
States of America 98:543-547 [arXiv:nlin.AO/0005039]

\bibitem{jeon2001}
Jeong, H., Mason S.~P.\ (2001) Lethality and Centrality in Protein 
Networks. Nature (London) 411:41-42 [arXiv:cond-mat/0105306]

\bibitem{jeon2000}
Jeong, H., Tombor, B., Albert, R., Oltvai, Z.~N., Barab\'asi, A.-L.\ (2000)
The Large-Scale Organization of Metabolic Networks. Nature (London) 
407:651-654 [arXiv:cond-mat/0010278]

\bibitem{jime1980}
Jim\'enez-Monta\~no, M.~A., Ebeling, W.\ (1980) A Stochastic 
Evolutionary Model of Technological Change. Collective 
Phenomena 3:107-114

\bibitem{kani2000}
Kaniovski Yu.~M.\ (2000) A Comparison of Diffusion Approximations
and Actual Limits in Births and Death Processes of Noisy Evolution.
Journal of Evolutionary Economics 10:545-555

\bibitem{katz1999}
Katz, J.~S.\ (1999) The Self-Similar Science System.
Research Policy  -- A Journal Devoted to Research Policy,
Research Management and Planning 28:501-517

\bibitem{kauf1995}
Kauffman, S.\ (1995) 
At Home in the Universe: The Search for Laws of Self-Organization 
and Complexity. Oxford University Press, New York

\bibitem{kirm2003}
Kirman, A.\ (2003) Economic Networks. In: Bornholdt, S., Schuster, H.~G.\
(Eds.) Handbook of Graphs and Networks: From the Genome to the Internet. 
Wiley-VCH Verlag, Weinheim, pp.\ 273-294 

\bibitem{klim1995}
Klimontovich, Yu.~L.\ (1995) Statistical Theory of Open Systems. 
Vol.\ 1: A Unified Approach to Kinetic Description of Processes 
in Active Systems. Fundamental Theories of Physics, Vol.\ 67.
Kluwer Academic Publishers, Dordrecht, Boston, London

\bibitem{kohl1926}
Kohlrausch, K.~W.~F., Schr\"odinger, E.\ (1926) Das Ehrenfestsche 
Modell der H-Kurve [Ehrenfest's model of the H-curve]. 
Physikalische Zeitschrift 27:306-313 [in German]. Reprinted
in: Schr\"odinger, E.\ (1984) Collected Papers/Gesammelte
Abhandlungen, Volume 1/Band 1, Contributions to Statistical 
Mechanics/Beitr\"age zur statistischen Mechanik. Verlag der 
\"Osterreichischen Akademie der Wissenschaften, Friedr.\ 
Vieweg \& Sohn, Braunschweig, Wiesbaden, pp.\ 349-357

\bibitem{kowo2003}
Kowol, U., K\"uppers, G.\ (2003). Innovation Networks: A New Approach 
to Innovation Dynamics. In: van Geenhuizen, M., Gibson, D.~V.,
Heitor M.~V.\ (Eds.) Regional Development and Conditions for 
Innovation in the Network Society. International Series on 
Technology Policy and Innovation.
Purdue University Press, Indiana, (in press)

\bibitem{krob1967}
Kr\"ober, G.\ (1967) Strukturgesetz und Gesetzesstruktur
[Law of Structure and Structure of Law].
Deutsche Zeitschrift f\"ur Philosophie 15:202-216 [in German]

\bibitem{laue1970}
Laue, R.\ (1970) Elemente der Graphentheorie und deren Anwendung 
in den biologischen Wissenschaften. Geest \& Portig, Leipzig;
(1971) Vieweg, Braunschweig [in German]

\bibitem{leyd2001}
Leydesdorff, L.\ (2001, 2.\ rev.\ ed. 2003) A Sociological 
Theory of Communication: The Self-Organization of the 
Knowledge-Based Society.
Universal Publishers/uPUBLISH.com, Parkland, FL

\bibitem{main1997phys}
Mainzer, K.\ (1997) Komplexe Systeme in Natur und Gesellschaft
[Complex Systems in Nature and Society].
Physik in unserer Zeit 28:74-81 [in German]

\bibitem{main1997think}
Mainzer, K.\ (1997) Thinking in Complexity: The Complex Dynamics of 
Matter, Mind, and Mankind, 3.\ rev.\ enl.\ ed.. 
Springer-Verlag, Berlin, Heidelberg, New York

\bibitem{main1999}
Mainzer, K.\ (Ed.) (1999)
Komplexe Systeme und Nichlineare Dynamik in Natur und Gesellschaft:
Komplexit\"atsforschung in Deutschland auf dem Weg ins n\"achste 
Jahrhundert. Springer-Verlag, Berlin, Heidelberg, New York [in German]

\bibitem{maur2000}
Maurer, S.~M., Huberman, B.~A.\ (2000) Competitive Dynamics of 
Web Sites. Journal of Economic Dynamics and Control 27:2195-2206
[arXiv:nlin.CD/0003041].
This article is part of the special issue: 
Juillard, M., Marcet, A.\ (Guest Eds.)
Computing in Economics and Finance. Proceedings of the 
6.\ International Conference of the Society for Computational
Economics, Barcelona, Spain, July 6-8, 2000.
Journal of Economic Dynamics and Control 27(11-12):1939-2265

\bibitem{milg1967}
Milgram, S.\ (1967) The Small World Problem.
Psychology Today 1:60-67

\bibitem{may1972}
May, R.~M.\ (1972) Will a Large Complex System Be Stable? 
Nature (London) 238:413-414

\bibitem{mole1962}
Moles, A.~A.\ (1962) Produkte: Ihre funktionelle und strukturelle 
Komplexit\"at [Products: Their Functional and Structural Complexity]. 
Ulm -- Zeitschrift der Hochschule f\"ur Gestaltung 6:4-12 [in German]

\bibitem{mont2002}
Montoya, J.~M., Sol\'e, R.~V.\ (2002) Small World Patterns in Food Webs.
Journal of Theoretical Biology 214:405-412 [arXiv:cond-mat/0011195]

\bibitem{more1978}
Moreno, J.~L.\ (1978) Who shall survive? Foundations of Sociometry, 
Group Psychotherapy and Sociodrama, 3.\ ed..
Beacon House, Beacon, NY

\bibitem{nach1998}
Nachtigall, C.\ (1998) Selbstorganisation und Gewalt. 
Internationale Hoch\-schulschriften,
Vol.\ 259. Waxmann Verlag, M\"unster, New York, M\"unchen, Berlin
[in German]

\bibitem{nagu2003}
Nagurney, A.\ (2003) Report: Some Recent Developments in Network Economics. 
Networks 41:68-72

\bibitem{newm1999}
Newman, M.~E.~J., Watts, D.~J.\ (1999) Scaling and Percolation in the
Small-World Network Model. Physical Review E 60:7332-7342
[arXiv:cond-mat/9904419]

\bibitem{newm2000}
Newman, M.~E.~J.\ (2000) Models of the Small World.
Journal of Statistical Physics 101:819-841
[arXiv:cond-mat/0001118]

\bibitem{newm2001}
Newman, M.~E.~J.\ (2001) Scientific Collaboration Networks: I. Network 
Construction and Fundamental Results. Physical Review E 64:016131, 8 pp.;
Scientific Collaboration Networks: II. Shortest Paths, Weighted Networks, 
and Centrality. Physical Review E 64:016132, 7 pp.
[arXiv:cond-mat/0011144]

\bibitem{nico1977}
Nicolis, G., Prigogine, I.\ (1977) Self-Organization in 
Non-Equilibrium Systems: From Dissipative Structures to Order 
Through Fluctuations. John Wiley \& Sons, New York 

\bibitem{prig1955}
Prigogine, I.\ (1955) Introduction to Thermodynamics of Irreversible 
Processes. 
American Lecture Series, Vol.\ 185. Charles C.\ Thomas Publisher, 
Springfield, IL.
(1962) 2.\ rev.\ ed., (1967) 3.\ ed.. Interscience Publishers, New York 

\bibitem{prig1987}
Prigogine, I., Sanglier, M.\ (Eds.) (1987) Laws of Nature and Human
Conduct: Specificities and Unifying Themes.
Proceedings of the 8.\ International Discoveries Symposium, 
Brussels, Belgium, October 7-9, 1985.
G.O.R.D.E.S./Task Force of Research, Information and Study of Science,
Brussels

\bibitem{prig1993}
Prigogine, I., Stengers, I.\ (1993) Das Paradox der Zeit: Zeit, Chaos 
und Quanten. Piper-Verlag, M\"unchen, Z\"urich [in German]

\bibitem{pyka1999}
Pyka, A.\ (1999) Der kollektive Innovationsproze\ss : Eine theoretische 
Ana\-ly\-se informeller Netzwerke und absorptiver F\"ahigkeiten.
Volkswirtschaftliche Schriften, Vol.\ 498. Duncker \& Humblot, Berlin
[in German]

\bibitem{pyka2002}
Pyka, A.\ K\"uppers, G.\ (Eds.) (2002) Innovation Networks: Theory 
and Practice. New Horizons in the Economics of Innovation Series. 
Edward Elgar Publishing, Cheltenham, Northampton, MA

\bibitem{rash1960}
Rashevsky, N.\ (1960) Mathematical Biophysics: Physico-Mathematical 
Foundations of Biology, 3.\ rev.\ ed., Vol.\ II. 
Dover Science Books, Vol.\ 575. Dover Publications, New York

\bibitem{rash1965}
Rashevsky, N.\ (1965) The Representation of Organisms in Terms of 
Predicates. The Bulletin of Mathematical Biophysics 27:477-492

\bibitem{rech1994}
Rechenberg, I.\ (1994) Evolutionsstrategie '94.
Werkstatt Bionik und Evolutionstechnik, Vol.\ 1. 
 Frommann-Holzboog-Verlag, Stuttgart, Bad Cannstatt [in German]

\bibitem{reny1959}
R\'enyi, A.\ (1959) Some Remarks on the Theory of Trees. 
A Magyar Tu\-do\-m\'anyos Akad\'emia Matematikai \'es Fizikai 
Tudom\'anyok Oszt\'aly\'anak K\"ozlem\'enyei    
[Publications of the Mathematical Institute of the Hungarian Academy of
Sciences] 4:73--85.
Reprinted in \cite{tura1976}, item 163, pp.\ 363-374

\bibitem{savi2000net}
Saviotti, P.~P.\ (2000) Networks, National Innovation Systems and 
Self-Organization. Paper presented at the 
4.\ International Conference on Technology Policy and Innovation --
Learning and Knowledge, Networks for Development,
CURITIBA 2000, Curitiba, Brazil, August 28-31, 2000.
Available from the conference web site:
http://in3.dem.ist.utl.pt/curitiba2000/default.htm
at the URL:
http://in3.dem.ist.utl.pt/downloads/cur2000/papers/S26P01.PDF

\bibitem{savi1996}
Saviotti, P.~P.\ (1996) Technological Evolution, Variety and the Economy.
Edward Elgar Publishing, Cheltenham, Northampton, MA

\bibitem{savi1995}
Saviotti, P.~P., Mani, G.~S.\ (1995) Competition, Variety and 
Technological Evolution: A Replicator Dynamics Model. 
Journal of Evolutionary Economics 5:369-392

\bibitem{savi2000noot}
Saviotti, P.~P., Nooteboom B.\ (Eds.) (2000) Technology and Knowledge: 
 From the Firm to Innovation Systems. Edward Elgar Publishing, Cheltenham,
Northampton, MA

\bibitem{scha1999}
Scharnhorst, A.\ (1999) Modelle von Wertedynamik und Kompetenzentwicklung 
[Models for the Dynamics of Values and Competence Development]. In:
Erpenbeck, J., Heyse, J.\ (Eds.) Kompetenzbiographie -- Kompetenzmileu 
-- Kompetenztransfer. QUEM-Report -- Schriften zur beruflichen Weiterbildung,
Issue 62. Arbeitsgemeinschaft Betriebliche Weiterbildungsforschung
e.V.\ (ABWF), Berlin, pp.\ 106-140 [in German]

\bibitem{scha2003}
Scharnhorst, A.\ (2003) Complex Networks and the Web: Insights from
Nonlinear Physics. Journal of Computer-Mediated Communication 8(4):\hfill

http://www.ascusc.org/jcmc/vol8/issue4/scharnhorst.html. 
This article is part of the special issue:
Beaulieu, A., Park, H.~W.\ (Guest Eds.) (2003) 
Internet Networks: The Form and the Feel.
Journal of Computer-Mediated Communication 8(4):
http://www.ascusc.org/jcmc/vol8/issue4

\bibitem{schi1981}
Schimansky-Geier, L.\ (1981)  Stochastische Theorie der 
Nichtgleichgewichts\-phasen\"uberg\"ange in einkomponentigen 
bistabilen chemischen Reaktionssystemen. Dissertation A (Ph.D.) Thesis,
Humboldt University Berlin, Berlin [in German]

\bibitem{schw1977}
Schwefel, H.-P.\ (1977) Numerische Optimierung von Computer-Modellen 
mittels der Evolutionsstrategie: Mit einer vergleichenden 
Einf\"uhrung in die Hill-Climbing- und Zufallsstrategie.
Interdisciplinary Systems Research, Vol.\ 26.
Birkh\"auser-Verlag, Basel, Stuttgart.
English translation: (1981) Numerical Optimization of Computer Models.
John Wiley \& Sons, Chichester, New York 

\bibitem{schw1997self}
Schweitzer, F.\ (1997). Self-Organization of Complex Structures:
 From Individual to Collective Dynamics. Gordon and Breach
Science Publishers, Amsterdam

\bibitem{schw1997}
Schweitzer, F., Ebeling, W., Ros\'e, H., Weiss, O.\ (1997) Optimization 
of Road Networks Using Evolutionary Strategies.
Evolutionary Computation 5:419-438

\bibitem{schw1997urb}
Schweitzer, F., Steinbrink, J.\ (1997) Urban Cluster Growth: Analysis 
and Computer Simulation of Urban Aggregations.
In: Schweitzer, F.\ (Ed.) Self-Organization of Complex Structures: 
 From Individual to Collective Dynamics.
Gordon and Breach Science Publishers, Amsterdam, pp.\ 501-518

\bibitem{scot2000}
Scott, J.\ (2000) Social Network Analysis: A Handbook, 2.\ ed..
SAGE Publications, London, Thousands Oaks, CA

\bibitem{silv2002discr}
Silverberg, G.\ (2002) The Discrete Charme of the Bourgeoisie: 
Quantum and Continuous Perspectives on Innovation and Growth. 
Research Policy -- A Journal Devoted to Research Policy,
Research Management and Planning: 31:1275-1289.
This article is part of the special issue: 
Freeman, Ch., Pavitt, K.\ (Guest Eds.)
NELSON + WINTER + 20. Special issue containing papers 
presented at the DRUID Nelson and Winter Conference,
Aalborg, Denmark, June 12-15, 2001.
Research Policy -- A Journal Devoted to Research Policy,
Research Management and Planning 31(8-9):1221-1528

\bibitem{silv2002perc}
Silverberg, G., Verspagen, B.\ (2002). A Percolation Model of 
Innovation in Complex Technology Spaces. Paper presented at the
8.\ International Conference of the Society for Computational Economics on 
`Computing in Economics and Finance', Aix-en-Provence, France,
June 27-29, 2002 (conference web site:
http://www.cepremap.cnrs.fr/sce2002.html).
Available at the `Computing in Economics and Finance 2002'
web site as paper no.\ 24 at the URL:
http://ideas.repec.org/p/sce/scecf2/24.html.
MERIT -- Infonomics Research Memoranda, No.\ 2002-025.
Maastricht Economic Reseach Institute on Innovation and Technology
(MERIT), Maastricht. Available at the URL:\hfill

http://www.merit.unimaas.nl/publications/rmpdf/2002/rm2002-025.pdf.

ECIS -- Eindhoven Centre for Innovation Studies, Working Paper No.\ 02.12.
Available at the URL:\hfill

http://www.tm.tue.nl/ecis/WorkingPapers/eciswp64.pdf

\bibitem{skyr2000}
Skyrms, B., Pemantle, R.\ (2000) A Dynamic Model of Social Network 
Formation. Proceedings of the National Academy of Sciences of the 
United States of America 97:9340-9346

\bibitem{sole2001}
Sol\'e, R.~V., Montoya, J.~M.\ (2001) Complexity and Fragility in 
Ecological Networks. Proceedings of the Royal
Society of London, Series B -- Biological Sciences 268:2039-2045
[arXiv:cond-mat/0011196]

\bibitem{sonn1984}
Sonntag, I.\ (1984) Random Networks of Catalytic Biochemical Reactions.
Biometrical Journal 26:799-807 

\bibitem{sonn1984app}
Sonntag, I.\ (1984) Application of the Percolation Theory to Random 
Networks of Biochemical Reactions. Biometrical Journal 26:809-813

\bibitem{sonn1981}
Sonntag, I., Feistel, R., Ebeling, W.\ (1981) Random Networks of 
Catalytic Biochemical Reactions. Biometrical Journal 23:501-515

\bibitem{stro2001}
Strogatz, S.~H.\ (2001) Exploring Complex Networks. Nature (London)
410:268-276

\bibitem{temk1996}
Temkin O.~N., Zeigarnik, A.~V., Bonchev, D.\ (1996) Chemical Reaction 
Networks: A Graph-Theoretical Approach. CRC Press, Boca Raton, FL

\bibitem{tura1976}
Tur\'an, P.\ (Ed.) (1976) R\'enyi Alfr\'ed V\'alogatott Munk\'ai/Selected 
Papers of Alfr\'ed R\'enyi, (Volume) 2, 1956-1961. Akad\'emiai Kiad\'o,
Budapest

\bibitem{tyle2003}
Tyler, J.~R., Wilkinson, D.~M., Huberman, B.~A.\  (2003) 
Email as Spec\-troscopy: Automated Discovery of Community Structure 
within Organizations [arXiv:cond-mat/0303264]. 
In: Huysman, M., Wenger, E., Wulf, V.\ (Eds.)
(2003) Communities and Technologies.
Proceedings of the First International Conference on Communities and
Technologies, C\&T 2003, September 19-21, 2003, Amsterdam, The Netherlands. 
Kluwer Academic Publishers, Dordrecht,
pp.\ 81-96

\bibitem{vazq2001}
Vazquez, A.\ (2001). Statistics of Citation Networks. Physics
E-Print Archive: arXiv:cond-mat/0105031
(http://arxiv.org/abs/cond-mat/0105031)

\bibitem{voge2000}
Vogelstein, B., Lane, D., Levine, A.~J.\ (2000) Surfing the p53 Network.
Nature (London) 408:307-310

\bibitem{wagn2000}
Wagner, A., Fell, D.\ (2000) The Small World Inside Large Metabolic 
Networks. Proceedings of the Royal Society of London, Series B -- 
Biological Sciences 268:1803-1810

\bibitem{wass1994}
Wasserman, S., Faust, K.\ (1994) Social Network Analysis: Methods 
and Applications. Structural Analysis in the Social Sciences, Vol.\ 8.
Cambridge University Press, Cambridge, New York

\bibitem{watt1999}
Watts, D.~J.\ (1999) Small Worlds: The Dynamics of Networks between 
Order and Randomness. Princeton Studies in Complexity.
Princeton University Press, Princeton, NJ

\bibitem{watt1998}
Watts, D.~J., Strogatz, S.~H.\ (1998) Collective Dynamics of 
`Small World' Networks. Nature (London) 393:440-442

\bibitem{weid2000}
Weidlich, W.\ (2000)  Sociodynamics: A Systematic Approach to 
Mathematical Modelling in the Social Sciences. 
Harwood Academic Publishers, Amsterdam

\bibitem{weid1983}
Weidlich, W., Haag, G.\ (1983) Concepts and Models of Quantitative 
Sociology: The Dynamics of Interacting Populations. 
Springer Series in Synergetics, Vol.\ 14. Springer-Verlag, 
Berlin, Heidelberg, New York

\bibitem{zima2000}
Ziman, J.\ (Ed.) (2000) Technological Innovation as an Evolutionary
Process. Cambridge University Press, Cambridge, New York

\end{thebibliography}
\end{document}